\documentclass[useAMS,usenatbib,usegraphicx]{mn2e}
\usepackage{subfig,amsmath,amssymb}

\def\mnras{MNRAS}
\def\apj{ApJ}

\def\aap{A\&A}

\def\aj{AJ}

\title[The micro-Jy source population at 8.4 GHz in the WHDF]{Sub-millimetre source identifications and the micro-Jansky source population at 8.4 GHz in the William Herschel Deep Field}
\author[Heywood et al.]{\parbox{\textwidth}{I.~Heywood$^{1,3}$\thanks{{\tt ianh@astro.ox.ac.uk}, SEPnet fellow},  R.~M.~Bielby$^{2}$, M.~D.~Hill$^{2}$, N.~Metcalfe$^{2}$, S.~Rawlings$^{1}$,T.~Shanks$^{2}$, and O.~M.~Smirnov$^{3,4}$}\vspace{0.4cm}\\
$^{1}$Astrophysics, Department of Physics, University of Oxford, Keble Road, Oxford, OX1 3RH, UK\\
$^{2}$Department of Physics, Durham University, South Road, Durham, DH1 3LE, UK\\
$^{3}$Department of Physics and Electronics, Rhodes University, PO Box 94, Grahamstown, 6140, South Africa\\
$^{4}$SKA South Africa, 3rd Floor, The Park, Park Road, Pinelands, 7405, South Africa}

\begin{document}

\date{Accepted 20XX Month XX. Received 20XX Month XX; in original form 20XX Month XX}

\pagerange{\pageref{firstpage}--\pageref{lastpage}} \pubyear{20XX}

\maketitle

\label{firstpage}

\begin{abstract}

Sub-millimetre observations of the William Herschel Deep Field (WHDF)
using the Large Apex Bolometer Camera (LABOCA) revealed possible sub-mm
counterparts for two X-ray absorbed quasars. The primary aim here is to
exploit Expanded Very Large Array (EVLA) radio continuum imaging at 8.4
GHz to establish the absorbed quasars as radio/sub-mm sources. The main
challenge in reducing the WHDF EVLA data was the presence of a strong 4C source
at the field edge. A new calibration algorithm was applied to the data to model and subtract this source.
The resulting thermal noise limited
radio map covers a sky area which includes the $16'\times16'$ Extended WHDF.
It contains 41 radio sources above the 4$\sigma$ detection threshold, 17
of which have primary beam corrected flux densities. The radio observations show that the two absorbed AGN 
with LABOCA detections are also coincident with radio sources,
confirming the tendency for X-ray absorbed AGN to be sub-mm bright.
These two sources also show strong ultraviolet excess (UVX) which
suggest the nuclear sightline is gas- but not dust-absorbed. Of the
three remaining LABOCA sources within the $\approx5'$ half-power
diameter of the EVLA primary beam, one is identified with a faint
nuclear X-ray/radio source in a nearby galaxy, one with a faint radio
source and one is unidentified in any other band.

More generally, differential radio source counts  calculated from the beam-corrected
data are in good agreement with previous observations, showing at
$S<50\mu$Jy a significant excess over a pure AGN model. In the full
area, of ten  sources fainter than this limit, six have optical
counterparts of which three are UVX (i.e.~likely quasars) including the
two absorbed quasar LABOCA sources. The other faint radio counterparts
are not UVX but are only slightly less blue and  likely to be
star-forming/merging galaxies, predominantly at lower luminosities and
redshifts. The four faint, optically unidentified radio sources may be
either dust obscured quasars or galaxies. These high redshift obscured
AGN and lower redshift star-forming populations are thus the main
candidates to explain the observed excess in the faint source counts and
hence also the excess radio background found previously by  the Absolute
Radiometer for Cosmology, Astrophysics and Diffuse Emission (ARCADE2)
experiment.
\end{abstract}


\section{Introduction}

The William Herschel Deep Field (WHDF; 7 $\times$ 7 arcminutes at 00h 22m 30s
+00$^{\circ}$ 21m 00s) has been comprehensively targeted by many authors
using many observatories for over two decades (Metcalfe et al.,~1991,
1995, 2001, 2006; McCracken et al.,~2000). It is one of the deepest
survey fields in terms of optical and near infrared photometry, having
more than 150 hours of CCD imaging on the 4.2~m William Herschel
Telescope (WHT) and 3.9~m telescope at Kitt Peak National Observatory,
resulting in magnitude limits of U~=~27, B~=~28.2, R~=~27 and
I~=~25.5. The magnitude apertures vary slightly
with band / seeing but are $\approx1.''3$ radius for the faintest stars.
All magnitudes quoted are in the Vega system. More details are given by
Metcalfe et al. (2001). There is also deep H band imaging following 30 hours of
exposure with the Calar Alto 3.5~m telescope (Vallbe-Mumbru, 2004). Deep,
high-resolution I-band imaging is available through Hubble Space
Telescope (HST) Advanced Camera for Surveys (ACS) imaging (Bohm \&
Zeigler, 2006). Observations with the Low Dispersion Survey Spectrograph
2 (LDSS-2) spectrograph on the 6.5~m Magellan telescope (Vallbe-Mumbru,
2004) and Very Large Telescope (VLT, 8.2~m) observations using FORS2
(Bohm \& Zeigler, 2006) have yielded $\sim$200 spectroscopic redshifts
in the field.

Moving away from optical and infrared wavelengths, the WHDF has deep
X-ray imaging via 75~ks of \emph{Chandra} ACIS-I data (Vallbe-Mumbru,
2004), and Bielby et al.~(2012) recently obtained sub-mm observations of
the field using the Large Apex Bolometer Camera (LABOCA), reaching a
depth of 1~mJy at a wavelength of 870~$\mu$m and resulting in the detection
of eleven sources.

This paper presents the latest addition to the multiwavelength coverage
of the WHDF, namely 8.4 GHz observations using the Expanded Very Large
Array (hereafter EVLA, since renamed the Karl G. Jansky Very Large
Array). These observations were motivated by the recent LABOCA data with
the high resolution afforded by the radio observations being required to
confirm whether any of the bright sub-mm sources in the field were
associated with the significant sample of both absorbed  and unabsorbed
quasars (QSOs) also present in the WHDF. In particular, Bielby et al.~(2012)
suggested two X-ray absorbed high redshift active galactic nuclei (AGN), one of which is a Type 2
quasar showing only narrow lines in the optical, may be associated with
LABOCA sources, in line with previous suggestions of a
connection between X-ray absorbed AGN and sub-mm emission (Gunn 1999;
Page et al.~2004; Hill \& Shanks 2011). It is therefore important to
establish the reality of the association between the sub-mm and these
absorbed AGN as far as possible, using these radio data.

The WHDF is hitherto largely unexplored at radio wavelengths due to the
presence of a strong 4C source 6 arcminutes south of the field centre,
however this has been cleanly subtracted from the observations described
in this paper using a new calibration algorithm. A description of the
observations and the calibration process follows in Section
\ref{sec:obs}. The radio map and source catalogue are presented in
Section \ref{sec:results}. Section \ref{sec:counterparts} matches the radio sources with counterparts at other wavebands. These results are discussed in Section
\ref{sec:discussion}, and concluding remarks are made in Section
\ref{sec:conclusions}.

\section{Observations and data reduction}
\label{sec:obs}

The WHDF was observed with the EVLA in its most compact D-configuration between 28 March 2010 and 15 June 2010. A total of 35 hours were awarded under the Open Shared Risk Observing program, of which 30 hours were observed. The observations were recorded in 25 independent scheduling blocks (SBs) with durations of either 1.5 or 3.5 hours, resulting in 25 sets of visibility data.

The X-band receivers were used and the correlator was configured to deliver two spectral windows centred at 8.396 and 8.524 GHz, each providing 64 $\times$ 2 MHz channels giving 256 MHz of contiguous frequency coverage. For this continuum experiment an 8:1 frequency averaging was applied to make the data volume more manageable. No time averaging was applied, so the standard 1 second integration time was preserved. Each of the 25 Measurement Sets were then split further in order to separate each of the two spectral windows, resulting in a total of 50 unique data sets to be independently edited and calibrated.

Editing of the data was performed using the PlotMS tool within the NRAO CASA\footnote{Common Astronomy Software Applications: {\tt http://casa.nrao.edu}} package. In general the data were free from corruptions save for the occasional bursts of radio frequency interference which were simple to locate and excise. Two SBs exhibited severe amplitude drifts on a majority of baselines and in both spectral windows and were discarded outright.

For flux calibration a single observation of J0137+3309 (3C48) was performed at the start of each SB, and the phase calibrator J0022+0014 (4C 00+02; 0.64 Jy at X-band) was observed for 40 seconds for every 9 minutes of target observation. 

Standard calibration procedures were followed using CASA. The flux scale for the amplitude calibrator was set against a model image as 3C48 is slightly resolved in X-band observations, and this source was also used to calibrate the bandpass. The per-antenna complex gain solutions were generated for the phase calibrator source and then interpolated across the target field. A Python script was constructed to apply this standard procedure to each of the 50 flagged Measurement Sets automatically. Plots of the gain solutions were generated and inspected in order to ensure successful calibration. Phase stability was excellent throughout. As an additional diagnostic, calibrated maps of the phase calibrator and target field were also generated by the script.

Inspection of the images of the target field immediately shows why the WHDF is a challenge for radio interferometer observations, as the phase calibrator is situated approximately 6 arcminutes south of the target field and completely dominates the radio emission in each map. Furthermore, as with many well-studied multiwavelength fields, the WHDF is situated close to the celestial equator (Declination +20 arcminutes). When observing the celestial equator with an interferometer the locus traced by a given baseline in the $uv$ plane as the Earth rotates exhibits a strong east-west bias. The Fourier transform of the $uv$ coverage determines the point-spread function (PSF, also known as the dirty beam) of the observation, thus a $uv$ plane sampling pattern that is dominated by east-west structure results in strong north-south sidelobes in the PSF. As can be seen in the upper panel of Figure \ref{fig:beforeandafter} the sidelobes associated with the phase calibrator completely cover the target field, and these effects must be mitigated in order to obtain a scientifically useful radio map.

\subsection{Subtracting the phase calibrator from the target field}
\label{sec:subtraction}

\begin{figure}
\vspace{140mm}
\nonumber
\centering
\includegraphics[width= \columnwidth]{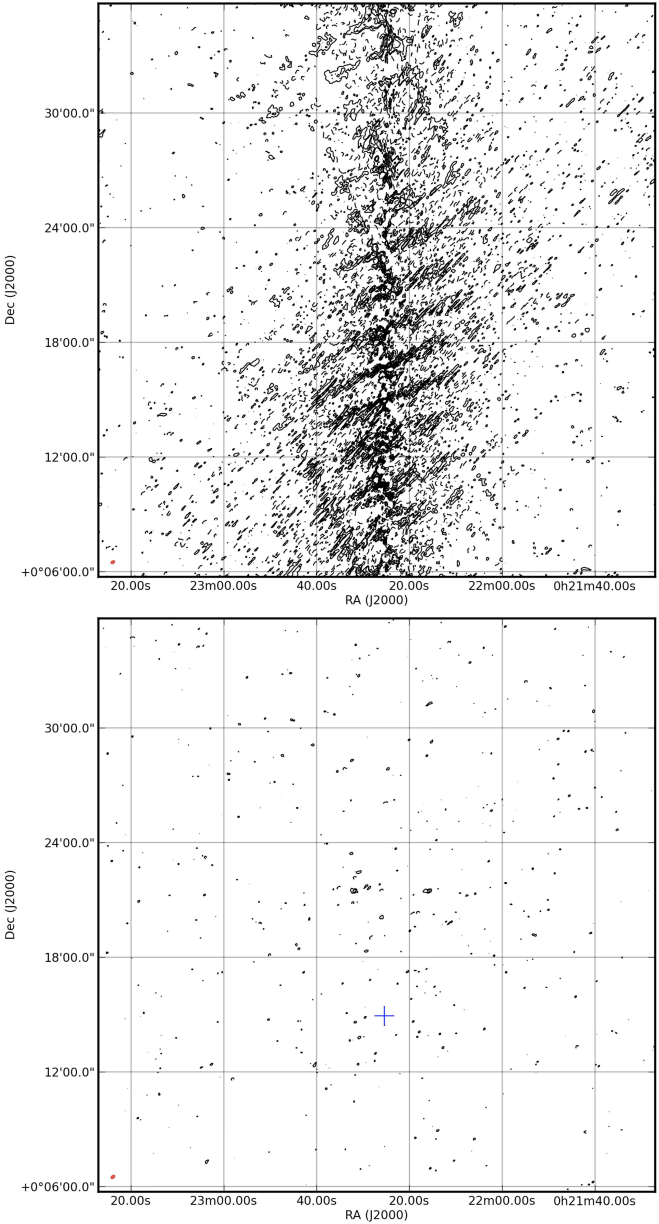} 
\caption{\label{fig:beforeandafter}Radio images of the target field generated from a single 3.5-hour Measurement Set. The upper panel shows the image that results from the
initial complex gain calibration and a multi-frequency synthesis deconvolution of the phase calibrator / confusing source using the CLEAN algorithm. 
The time-dependent gains imparted by the primary beam of the array in that direction corrupt the PSF associated with this source and even following deconvolution
the residual sidelobe structure completely covers the science targets. The lower panel shows the map that results following subtraction of the confusing 
source using the differential gains algorithm (Smirnov, 2011b). The position of the confusing source is marked with the cross. Radio sources in the target field are now visible. The contour levels in these maps are the same, and begin
at 3$\sigma$ (where $\sigma$~=~12 $\mu$Jy, the background rms of the lower map) and increase in multiples of 2. A single, dashed negative countour is 
included with a value of -3$\sigma$.}  
\end{figure}

Straightforward multifrequency synthesis deconvolution of the phase calibrator from the target field using the CLEAN algorithm did not yield the result that one might hope for, and residual sidelobe structure still swamped the target field: the upper panel of Figure 
\ref{fig:beforeandafter} actually shows the calibrated target field after deconvolution has been attempted. Similarly, subtraction of a visibility model derived from the clean components followed by imaging of the residual data set resulted in a map that was still corrupted by strong north-south emission associated with the phase calibrator.

Examining a model of the EVLA primary beam at 8.4 GHz (Walter Brisken, private communication; Figure \ref{fig:beam_patterns}) reveals that when the array is pointing at the target field the phase calibrator lies at the boundary of the first null and the first sidelobe, with its radial separation from the pointing centre changing with frequency. The first sidelobe of the EVLA beam has four-way azimuthal structure caused by the Cassegrain optics of the dishes, and the azimuth-elevation mount of the EVLA dishes causes this pattern to rotate on the sky as a source is tracked. Since the primary beam can to first order be thought of as the gain of the instrument as a function of direction, these effects conspire to impart apparent temporal variability to the confusing source (as well as significant spectral effects). Sources which vary significantly on timescales shorter than the observation have corrupted PSFs associated with them, and this is why traditional deconvolution and model subtraction cannot cope.

A calibration scheme which can solve for these direction-dependent effects must be employed, and for these observations the differential gains algorithm\footnote{Also known as ``The Flyswatter" for reasons that will become obvious.} (Smirnov, 2011b) was used, implemented using the Calico framework within the MeqTrees\footnote{{\tt http://www.astron.nl/meqwiki}} software package (Noordam \& Smirnov, 2011). The algorithm works by solving for additional complex gain terms against an assumed sky model for a subset of sources. In the case of these observations the solutions are generated for a sky model consisting solely of a point source at the location of the phase calibrator. Since this source dominates the radio emission, the gain solutions will also be dominated by corruptions in the direction of that source. There is no need to provide a stringent measure of the source flux for the sky model, as any errors in this parameter will be subsumed into the gain solutions, however the flux in the sky model was fixed at 5 mJy, as measured from the image generated following the initial calibration pass. Note the high attenuation of this source by comparing this measured value to its intrinsic flux measured by imaging the phase calibration scans. 

Data are averaged over five minute intervals and in pairs of frequency channels (i.e.~four blocks across the averaged band). A solution is generated for each of these time-frequency tiles. This has the effect of reducing the degrees of freedom in the fit, whilst simultaneously time-smearing out contributions from the rest of the field as well as accounting for any spectral behaviour in either the direction-dependent gain corruptions or the source itself. In this regime, the use of the differential gains algorithm is analogous to the more generally employed ``peeling" algorithm (e.g.~Noordam, 2004), however the advantage of the differential gains technique for more general applications is one of flexibility: it can generate additional gain terms for many individual sources simultaneously, whilst simultaneously solving on shorter timescales for the traditional complex receiver gains derived from an all-inclusive sky model. Peeling generally requires a cumbersome iterative approach whereby dominating sources are treated in order of brightness. 

The best fit visibility model derived from this process was subtracted from the observed visibilities, and the residual data were imaged. Following this procedure the confusing source is completely removed leaving no residual emission above the noise. As a quantitative example of the success of this of this technique, the root-mean-square (rms) background level in a map formed from one of the spectral windows of one of the 3.5-hour scheduling blocks was 21 $\mu$Jy / beam (as measured away from the dominating residual north-south structure) following either deconvolution of the confusing source, or subtraction of an inverted clean component model from the visibilities. This value dropped to 12 $\mu$Jy / beam following the subtraction of the confusing source using the differential gains algorithm. The radio images corresponding to these ``before" and ``after" scenarios can be seen in Figure \ref{fig:beforeandafter} with further details provided in the caption.

% arXiv
%\begin{figure*}
%\begin{center}
%\setlength{\unitlength}{1cm}
%\begin{picture}(16,11.0)
%\put(-0.5,0.0){\special{psfile=before_and_after_compressed.eps vscale=90 hscale=90 voffset=0
%hoffset=0}}
%\end{picture}
%\caption{\label{fig:beforeandafter}Radio images of the target field generated from a single 3.5-hour Measurement Set. The upper panel shows the image that results from the
%initial complex gain calibration and a multi-frequency synthesis deconvolution of the phase calibrator / confusing source using the CLEAN algorithm. 
%The time-dependent gains imparted by the primary beam of the array in that direction corrupt the PSF associated with this source and even following deconvolution
%the residual sidelobe structure completely covers the science targets. The lower panel shows the map that results following subtraction of the confusing 
%source using the differential gains algorithm (Smirnov, 2011b). The position of the confusing source is marked with the cross. Radio sources in the target field are now visible. The contour levels in these maps are the same, and begin
%at 3$\sigma$ (where $\sigma$~=~12 $\mu$Jy, the background rms of the lower map) and increase in multiples of 2. A single, dashed negative countour is 
%included with a value of -3$\sigma$.}  
%\end{center}
%\end{figure*}

Following the discarding of the two corrupted SBs, the first step in the calibration produced 46 Measurement Sets which were then split in order to contain only calibrated visibilities from the target field. MeqTrees can be operated non-interactively, allowing the source subtraction process to be scripted and applied automatically to each of these Measurement Sets. Again, calibrated maps were produced for each individual Measurement Set and the differential gain solutions were examined to check for problems. The gain solutions themselves encode information as to the origin of the direction-dependent corruptions, and these are discussed in the Appendix.

\subsection{Concatenation and imaging}

The calibrated data with the confusing source removed were concatenated into a final Measurement Set which was imaged to form the final map. At this stage, only 23 of the 46 viable Measurement Sets were included. The ones which were not included were shorter duration runs at low elevations. The observations had relaxed hour angle constraints when they were scheduled, and the inclusion of these data did not improve the quality or depth of the final map, which is presented in Section \ref{sec:radiomap}.

\section{Results}
\label{sec:results}

The following two sections present the final radio map of the WHDF at 8.4 GHz and the source catalogue derived from it.

\subsection{Radio map}
\label{sec:radiomap}

\begin{figure*}
\begin{center}
\setlength{\unitlength}{1cm}
\begin{picture}(16,17)
\put(-1.1,-0.5){\includegraphics{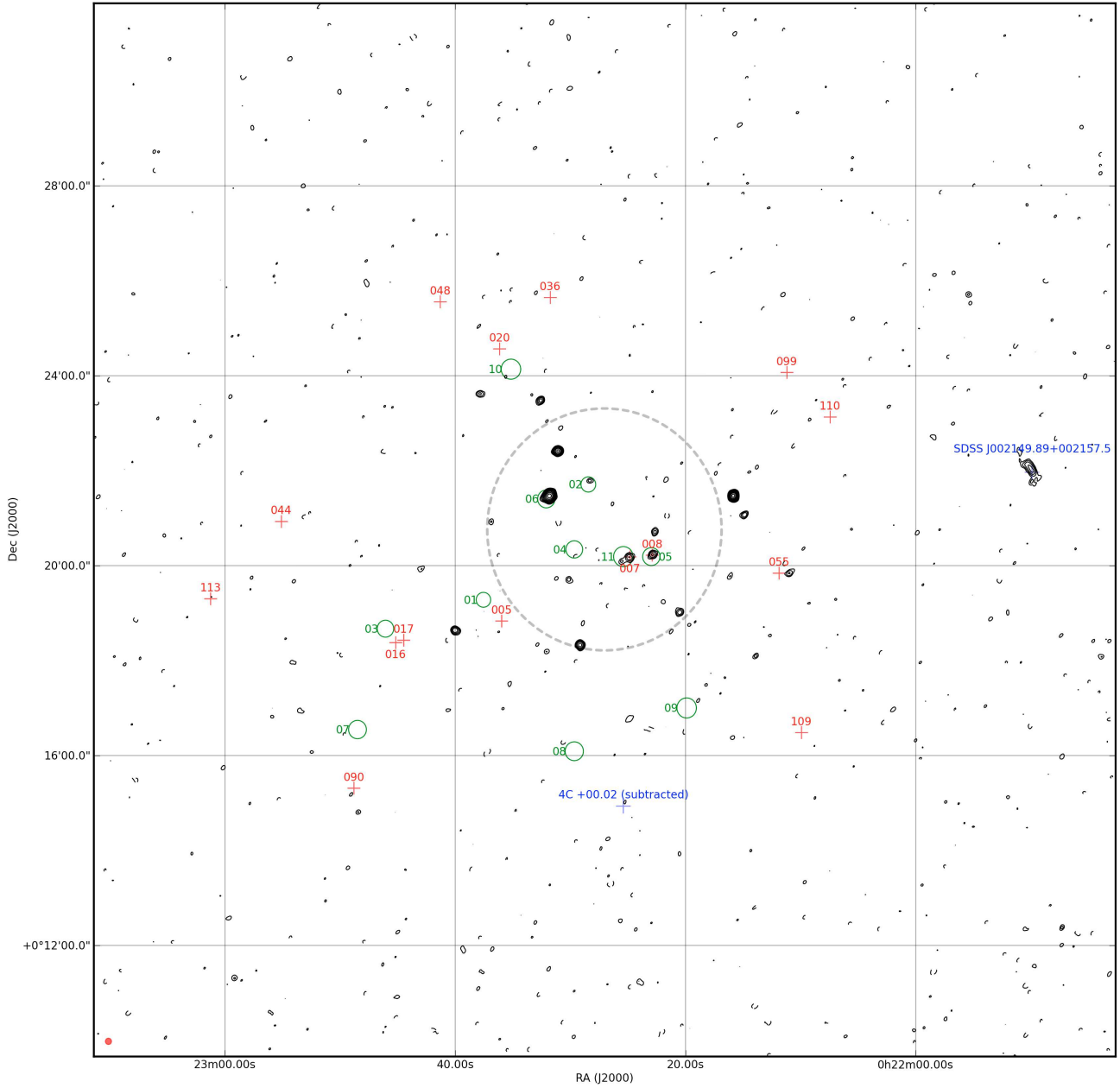}}
\end{picture}
\caption{\label{fig:radiomap} Radio contour map of the William Herschel
Deep Field at 8.4 GHz. The rms noise ($\sigma$) in the map is 2.5
$\mu$Jy / beam. Contours begin at 3$\sigma$ and increase in multiples of
$\sqrt{2}$. There is a single dashed negative contour at -3$\sigma$. The
numbered crosses show the spectroscopically confirmed quasars in the
WHDF (Vallbe-Mumbru, 2004) and the numbered circles indicate the
locations of the LABOCA-detected sub-mm sources (Bielby et al.,~2012).
The numbers associated with these two classes of sources are consistent
with those used by Bielby et al.~(2012), see Section \ref{sec:radiomap}
for details. The size of the circles used to show the locations of the
LABOCA sources indicates the positional uncertainty. The large dashed
circle shows the approximate half-power point of the EVLA primary beam.
The position of the phase calibrator 4C 00+02 is also marked although no
detectable emission remains. Note also the extended SDSS source to the
west. The filled circle in the lower left-hand corner of the plot shows
the 8 arcsecond extent of the circular restoring beam applied following
deconvolution.}
\end{center}
\end{figure*}

Figure \ref{fig:radiomap} shows the final, deconvolved 8.4 GHz radio image covering the WHDF. Deconvolution was performed using the CASA clean task in multi-frequency synthesis mode with w-term correction. The restoring beam is a circular Gaussian with a full-width at half-maximum spanning 8 arcseconds, as shown by the filled circle in the lower left-hand corner of the map. The rms of the background noise level in this map is 2.5 $\mu$Jy. This is consistent with the theoretical thermal noise level expected in this observation to within 3\%. The contours on Figure \ref{fig:radiomap} begin at 3$\sigma$ and increase in multiples of $\sqrt{2}$. There is a single dashed negative contour at -3$\sigma$.

The area of sky covered by this image is 22.2 $\times$ 22.2 arcminutes, although most of the multiwavelength observations occupy a 7 $\times$ 7 arcminute patch near the map centre. This is well-matched to the primary beam of the EVLA antennas, the approximate half-power point of which is shown by the large dashed circle in Figure \ref{fig:radiomap}. Imaging over this extended area in the radio was done to search for bright radio sources in the 16 $\times$ 16 arcminute area covered by the shallower Extended WHDF observations. The position of the phase calibrator source (4C 00+02) that was confusing the central region before subtraction is also marked. No residual emission is present above the noise. Also noteworthy is the spatially extended radio source towards the western edge of the map (WHDF-EVLA-3 in Table \ref{tab:sources}). This source is present in the Sloan Digital Sky Survey (SDSS; Abazajian et al.,~2009) which lists it as a galaxy at $z$~=~0.2609.

\begin{figure}
\vspace{70mm}
\nonumber
\centering
\includegraphics[width= \columnwidth]{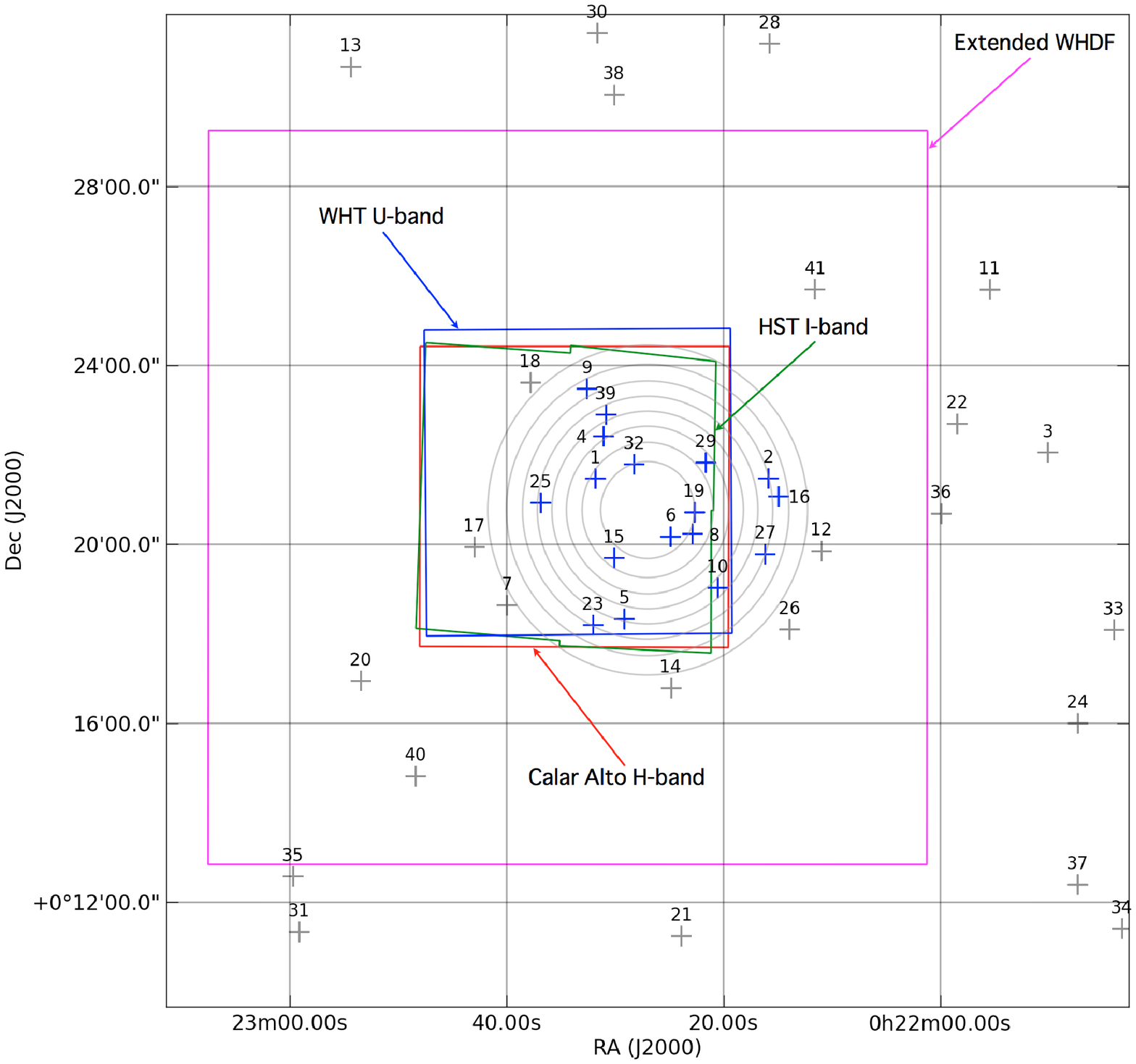} 
\caption{\label{fig:radiocat}Positions and IDs of the radio sources listed in Table \ref{tab:sources} and the coverage of the optical and infrared frames from which the thumbnails in Figure \ref{fig:evla_optical} have been extracted. The concentric grey contours show the main lobe of the EVLA primary beam. The outer contour is the 20\% gain level and the contours have increments of 10\%. Also shown is the area covered by the extended WHDF observations.}
\end{figure}

The crosses on Figure \ref{fig:radiomap} show the positions of the spectroscopically-confirmed quasars in the WHDF (Vallbe-Mumbru, 2004). The small circles show the locations of the LABOCA-detected sub-mm sources in the field, with the size of the circles being the positional uncertainties in the detections. The nomenclature of these two classes of sources on Figure \ref{fig:radiomap} and in the discussion that follows is a concise version of that employed by Bielby et al.~(2012): three digit identifications are confirmed quasars and two digit identifications correspond to LABOCA sources (prefixed by WHDFCH and WHDF-LAB respectively in the aforementioned article). Radio sources are referred to using the full WHDF-EVLA prefix throughout, with the exception of the labels on Figure \ref{fig:radiocat}. This covers the same sky area as Figure \ref{fig:radiomap} but shows the location of each source according to its identification as listed in Table \ref{tab:sources}. Figure \ref{fig:radiocat} also shows the central and extended regions of the WHDF. Higher magnification thumbnails of individual radio sources can be found in Figures \ref{fig:evla_optical} and \ref{fig:evla_optical_ext}, along with images of the counterparts identified at other wavebands (see Section \ref{sec:optical}).

\subsection{Source catalogue}
\label{sec:catalogue}

The radio map presented in Section \ref{sec:radiomap} was searched for
peaks of emission exceeding 4$\sigma$ (=~10 $\mu$Jy) using the SAD task
within the Astronomical Image Processing System (AIPS; Greisen, 2003)
package. The resulting catalogue was manually pruned as two of the
sources (WHDF-EVLA-3 and WHDF-EVLA-6) were resolved into multiple
Gaussian components. In these cases the flux was determined by summing
up the components. The 22.2~$\times$~22.2 arcminute map shown in Figure
\ref{fig:radiomap} contains 41 discrete sources with flux densities
exceeding 4$\sigma$ and the properties of these are listed in descending
order of brightness in Table \ref{tab:sources}.

Of the sources detected in the map, 17 of them were within the region
where the gain of the main lobe of the EVLA primary beam exceeded 0.2.
These are indicated by the circles in Table \ref{tab:sources} (see also
Figure \ref{fig:radiocat}). An analytic expression is available to
correct for the beam gain of the (E)VLA antennas as per the AIPS task
PBCOR, and this has been applied to the 17 sources in the main lobe.
Thus bold-face type in Table \ref{tab:sources} represents intrinsic flux
values and plain type entries have apparent flux values. The measured
separation from the pointing centre that was used to derive the primary
beam correction factor for each source is also listed. Uncertainties in
the intrinsic fluxes include contributions from errors in the fit to the
main lobe of the beam.

The subtracted phase calibrator (0.64 Jy intrinsic flux) is not included in this
table. If one were interested in this source it would be more useful to
simply image the phase calibrator scans during which the array was
pointing directly at it rather than trying to correct for its
complicated behaviour in the target scans.

\begin{table*}
\begin{minipage}{170mm}
\centering
\begin{tabular}{lllllllll} \hline
& ID             & Right Ascension & Declination     & ($\sigma_{RA},\sigma_{Dec}$) & Flux density & Radius   & Beam-corrected  \\ 
&                & (J2000)         & (J2000)         & (arcsec)                     & ($\mu$Jy)    & (arcmin) & flux density ($\mu$Jy)     \\\hline		
$\circ$ & WHDF-EVLA-1 & 0h 22m 31.82s & +0d 21m 28.03s  & (0.1,0.1) & 164.42 ($\pm$6.0) &   1.389 & {\bf 201.23 ($\pm$7.55)}  \\
$\circ$ & WHDF-EVLA-2 & 0h 22m 15.86s & +0d 21m 28.32s  & (0.08,0.09) & 106.23 ($\pm$4.0) &   2.884 & {\bf 275.71 ($\pm$18.4)}  \\
        & WHDF-EVLA-3 & 0h 21m 50.09s & +0d 22m 3.15s   & (0.64,0.77) & 87.58 ($\pm$10.0) &   9.33 & N/A   \\
$\circ$ & WHDF-EVLA-4 & 0h 22m 31.07s & +0d 22m 24.81s  & (0.13,0.12) & 70.88 ($\pm$4.0) &   1.939 & {\bf 106.16 ($\pm$6.84)}  \\
$\circ$ & WHDF-EVLA-5 & 0h 22m 29.15s & +0d 18m 19.78s  & (0.24,0.26) & 43.85 ($\pm$5.0) &   2.482 & {\bf 86.98 ($\pm$9.76)}  \\
$\circ$ & WHDF-EVLA-6 & 0h 22m 24.89s & +0d 20m 10.21s  & (0.68,0.61) & 43.1 ($\pm$7.0) &   0.794 & {\bf 45.99 ($\pm$7.81)}  \\
& WHDF-EVLA-7 & 0h 22m 39.97s & +0d 18m 38.2s  & (0.31,0.28) & 33.38 ($\pm$5.0) &   3.867 & N/A   \\
$\circ$ & WHDF-EVLA-8 & 0h 22m 22.85s & +0d 20m 13.98s  & (0.48,0.43) & 32.51 ($\pm$5.0) &   1.169 & {\bf 37.47 ($\pm$6.28)}  \\
$\circ$ & WHDF-EVLA-9 & 0h 22m 32.6s & +0d 23m 28.48s  & (0.36,0.36) & 29.11 ($\pm$5.0) &   3.053 & {\bf 86.04 ($\pm$14.85)}  \\
$\circ$ & WHDF-EVLA-10 & 0h 22m 20.53s & +0d 19m 1.52s  & (0.55,0.56) & 28.3 ($\pm$6.0) &   2.375 & {\bf 52.75 ($\pm$10.6)}  \\
& WHDF-EVLA-11 & 0h 21m 55.42s & +0d 25m 41.58s  & (1.37,1.3) & 26.01 ($\pm$8.0) &   9.32 & N/A   \\
& WHDF-EVLA-12 & 0h 22m 10.95s & +0d 19m 50.99s  & (0.89,0.66) & 25.52 ($\pm$6.0) &   4.122 & N/A   \\
& WHDF-EVLA-13 & 0h 22m 54.37s & +0d 30m 39.98s  & (1.85,1.59) & 24.75 ($\pm$9.0) &   12.036 & N/A   \\
& WHDF-EVLA-14 & 0h 22m 24.83s & +0d 16m 47.18s  & (1.37,1.25) & 24.66 ($\pm$7.0) &   4.008 & N/A   \\
$\circ$ & WHDF-EVLA-15 & 0h 22m 30.09s & +0d 19m 41.63s  & (1.04,0.94) & 23.09 ($\pm$7.0) &   1.31 & {\bf 27.62 ($\pm$8.02)}  \\
$\circ$ & WHDF-EVLA-16 & 0h 22m 14.93s & +0d 21m 4.12s  & (0.5,0.43) & 22.97 ($\pm$5.0) &   3.046 & {\bf 67.52 ($\pm$14.43)}  \\
& WHDF-EVLA-17 & 0h 22m 42.96s & +0d 19m 56.28s  & (1.35,1.01) & 21.34 ($\pm$7.0) &   4.063 & N/A   \\
& WHDF-EVLA-18 & 0h 22m 37.79s & +0d 23m 36.78s  & (0.67,0.46) & 21.15 ($\pm$5.0) &   3.922 & N/A   \\
$\circ$ & WHDF-EVLA-19 & 0h 22m 22.66s & +0d 20m 42.82s  & (0.46,0.59) & 20.72 ($\pm$5.0) &   1.093 & {\bf 23.45 ($\pm$5.46)}  \\
& WHDF-EVLA-20 & 0h 22m 53.44s & +0d 16m 57.03s  & (1.48,1.04) & 20.17 ($\pm$7.0) &   7.618 & N/A   \\
& WHDF-EVLA-21 & 0h 22m 23.89s & +0d 11m 14.61s  & (1.04,1.21) & 20.12 ($\pm$7.0) &   9.547 & N/A   \\
& WHDF-EVLA-22 & 0h 21m 58.45s & +0d 22m 41.49s  & (1.26,1.17) & 18.88 ($\pm$7.0) &   7.405 & N/A   \\
$\circ$ & WHDF-EVLA-23 & 0h 22m 32.02s & +0d 18m 11.78s  & (1.51,1.02) & 18.67 ($\pm$7.0) &   2.846 & {\bf 47.13 ($\pm$17.25)}  \\
& WHDF-EVLA-24 & 0h 21m 47.32s & +0d 15m 59.91s  & (1.19,1.33) & 17.94 ($\pm$7.0) &   11.011 & N/A   \\
$\circ$ & WHDF-EVLA-25 & 0h 22m 36.86s & +0d 20m 55.57s  & (0.86,1.1) & 16.43 ($\pm$6.0) &   2.461 & {\bf 32.19 ($\pm$11.18)}  \\
& WHDF-EVLA-26 & 0h 22m 13.91s & +0d 18m 5.77s  & (0.98,0.82) & 15.95 ($\pm$5.0) &   4.225 & N/A   \\
$\circ$ & WHDF-EVLA-27 & 0h 22m 16.16s & +0d 19m 46.78s  & (0.97,1.04) & 15.01 ($\pm$5.0) &   2.891 & {\bf 39.16 ($\pm$13.9)}  \\
& WHDF-EVLA-28 & 0h 22m 15.75s & +0d 31m 11.39s  & (1.31,1.23) & 15.0 ($\pm$6.0) &   10.808 & N/A   \\
$\circ$ & WHDF-EVLA-29 & 0h 22m 21.64s & +0d 21m 49.48s  & (0.87,1.02) & 13.8 ($\pm$5.0) &   1.722 & {\bf 18.91 ($\pm$7.06)}  \\
& WHDF-EVLA-30 & 0h 22m 31.64s & +0d 31m 25.44s  & (0.8,1.24) & 12.99 ($\pm$5.0) &   10.729 & N/A   \\
& WHDF-EVLA-31 & 0h 22m 59.13s & +0d 11m 20.34s  & (0.68,0.75) & 12.95 ($\pm$4.0) &   12.372 & N/A   \\
$\circ$ & WHDF-EVLA-32 & 0h 22m 28.25s & +0d 21m 47.24s  & (0.79,0.54) & 12.94 ($\pm$4.0) &   1.073 & {\bf 14.58 ($\pm$4.88)}  \\
& WHDF-EVLA-33 & 0h 21m 43.99s & +0d 18m 5.18s  & (0.79,0.73) & 12.11 ($\pm$4.0) &   11.09 & N/A   \\
& WHDF-EVLA-34 & 0h 21m 43.24s & +0d 11m 24.65s  & (1.07,1.26) & 11.96 ($\pm$5.0) &   14.397 & N/A   \\
& WHDF-EVLA-35 & 0h 22m 59.7s & +0d 12m 34.81s  & (1.21,0.86) & 11.93 ($\pm$5.0) &   11.554 & N/A   \\
& WHDF-EVLA-36 & 0h 21m 59.92s & +0d 20m 41.05s  & (0.96,1.01) & 11.09 ($\pm$5.0) &   6.779 & N/A   \\
& WHDF-EVLA-37 & 0h 21m 47.34s & +0d 12m 23.71s  & (0.54,0.76) & 10.71 ($\pm$4.0) &   12.977 & N/A   \\
& WHDF-EVLA-38 & 0h 22m 30.07s & +0d 30m 2.58s  & (0.91,0.64) & 10.3 ($\pm$4.0) &   9.316 & N/A   \\
$\circ$ & WHDF-EVLA-39 & 0h 22m 30.84s & +0d 22m 53.9s  & (0.8,0.79) & 10.08 ($\pm$4.0) &   2.345 & {\bf 18.46 ($\pm$7.65)}  \\
& WHDF-EVLA-40 & 0h 22m 48.38s & +0d 14m 49.05s  & (0.72,0.56) & 9.36 ($\pm$4.0) &   7.982 & N/A   \\
& WHDF-EVLA-41 & 0h 22m 11.57s & +0d 25m 41.96s  & (0.68,0.67) & 9.27 ($\pm$4.0) &   6.277 & N/A   \\
\hline
\end{tabular}
\caption{Properties of the 41 radio sources detected in the EVLA map centred on the WHDF and presented in Figure \ref{fig:radiomap}. 
Sources marked with a circle ($\circ$) in the left-hand
column are within the region of the main lobe of the EVLA primary beam where the direction dependent gain is greater than 0.2. These
are corrected with a primary beam model, thus the flux density values in bold face type represent \emph{intrinsic} values, whereas
the ones in plain type are \emph{apparent} values. Please refer to the text for further details.}\label{tab:sources}
\end{minipage}
\end{table*}

\section{Sub-mm and optical/near-infrared radio source counterparts}
\label{sec:counterparts}

\subsection{Sub-mm counterparts}

High redshift extragalactic sources which are luminous at sub-mm wavelengths are
generally thought to be driven by high star formation rates, with the
high sub-mm flux in the system arising due to the reprocessing of the
intense stellar radiation by large amounts of dust. The  low temperature
of the dust  is thought to favour such a heating  source rather than AGN.
However the high redshift sub-mm sources clearly have the bolometric
luminosities of AGN. Moreover, explaining the observed sub-mm source counts as
being of purely star-formation origin in a Cold Dark Matter cosmology context requires
invocation of top heavy initial mass functions \citep{baugh2005}. 
Recently, evidence has mounted for AGN making significant contributions to  sub-mm  source counts.
(e.g.~\citealt{lutz2010,hill2011a}) following earlier identifications of  AGN in 
sub-mm sources (eg Ivison et al. 1998,  Brandt et al. 2001)

These findings motivated the 870~$\mu$m LABOCA observations of Bielby et
al.~(2012) as the WHDF conveniently contains 15 spectroscopically
confirmed quasars, four of which are classified as obscured via the
assumption that their high X-ray hardness ratio\footnote{If $H$ and $S$
represent photon counts in the hard (2--8 keV) and soft (0.5--2 keV)
X-ray bands then the hardness ratio is defined as ($H$~-~$S$) /
($H$~+~$S$).} is caused by the presence of large amounts of absorbing
hydrogen. Eleven sub-mm sources were detected in the LABOCA
observations, two of which appeared to be associated with heavily
obscured quasars. None of the unobscured quasars were near to robust
sub-mm counterparts, although a stacking analysis revealed that 
absorbed X-ray quasars showed the most significant sub-mm emission.

Since sub-mm observations of this kind have relatively low resolution,
radio observations have been used to detect counterparts for sub-mm
sources by several authors (e.g.~Ivison et al.,~2002), with the improved
positional accuracy afforded by the radio observations then being used
to confirm counterparts at other wavebands. Recalling the naming scheme
defined in Section \ref{sec:radiomap}, the EVLA observations detect
radio counterparts at the positions of four sub-mm sources: 06, 11, 05
and 02 corresponding to radio sources WHDF-EVLA-1, WHDF-EVLA-6,
WHDF-EVLA-8 and WHDF-EVLA-32. Radio sources WHDF-EVLA-6 and WHDF-EVLA-8
correspond to the X-ray absorbed quasars 007 ($z$~=~1.33) and 008 ($z$~=~2.12)
respectively.

The reliability of the radio and sub-mm sources being counterparts (and
not chance alignments) is evaluated here using the corrected Poisson
probability ($P$) as used by \citet{downes1986}. Based on the number
densities of radio sources at the fluxes of those presented here, the
probabilities that radio sources WHDF-EVLA-1,WHDF-EVLA-6, WHDF-EVLA-8,
and WHDF-EVLA-32 are chance alignments with the LABOCA sub-mm sources
are $P=0.003$, $P=0.023$, $P=0.006$ and $P=0.032$ respectively. These
probabilities of $\lesssim3\%$ indicate that the radio and sub-mm
signals do indeed originate from the same sources.

Summarising, of the 5 sub-mm sources in the central WHDF area surveyed
by EVLA, 4 are radio sources and 3 were already known to be candidate
X-ray sources. The EVLA observations confirm the positional coincidence
of the three sub-mm source with X-ray sources on the assumption that the
LABOCA and EVLA sources are the same. One further LABOCA source is close
to an EVLA source (02/WHDF-EVLA-32) but neither source is close to an
optical/near-infrared object. The final sub-mm source (04) is
unidentified in either radio or optical/near-infrared. Thus three out of
five sub-mm sources are AGN as indicated by their X-ray properties and
the other two are blank in the optical/near-infrared. Two of the three
X-ray sources are hard X-ray sources and likely to be cold gas absorbed.
Although the numbers of sub-mm sources are small, the fraction that are
X-ray sources is high and so is the gas-absorbed fraction. 

\citet{hill2011a}, following \citet{gunn1999}, suggested that the FIR flux re-radiated by the dust might
be proportional to the amount of X-ray radiation absorbed and in this
case the obscured AGN contribution to the sub-mm background might reach
40\%. \citet{hill2011a} found strong evidence for this through statistical analyses
of the Extended Chandra Deep Field South (ECDFS) X-ray and sub-mm data. Aided by the EVLA confirmations of
the identification of the 2 absorbed X-ray quasars with the sub-mm sources
means that there is now further evidence for this hypothesis in the new
WHDF/EVLA dataset. These points will be revisited in the  discussion in Section 
\ref{sec:discussion}.

\subsection{Optical/near-infrared counterparts and Individual radio sources}
\label{sec:optical}

\begin{table*}
\begin{minipage}{170mm}
\centering
\begin{tabular}{llllllllllll} \hline
Radio ID        & R-band RA & R-band Dec & Offset & R  & U-B & B-R & R-I & R-Z & R-H & H-K & Redshift  \\
                & (J2000)  & (J2000)& ($''$) & mag & \multicolumn{6}{|c|}{Colours within $3''$ diameter apertures} &  \\ 
\hline

WHDF-EVLA-1 & 0h 22m 31.86s & +0d 21m 27.5s  & 0.8 & 16.07 & 0.04 & 1.42 & 0.73 & 0.89 & 2.49 & 0.32 &0.046   \\
WHDF-EVLA-2 & 0h 22m 15.87s & +0d 21m 28.8s  & 0.5 & 19.20 & 0.52 & 2.70 & 0.92 & 1.11 & 3.04 &  -  &0.259   \\
WHDF-EVLA-4 & 0h 22m 31.04s & +0d 22m 24.7s  & 0.5 & 23.29 & 0.32 & 2.23 & 1.64 & 1.82 & 4.62 & 1.22  &0.961   \\
WHDF-EVLA-5 & 0h 22m 29.12s & +0d 18m 19.2s  & 0.7 & 21.62 & -0.12  &  2.01  & 0.97 & 1.32 & 3.59 & 0.79  &0.433   \\
WHDF-EVLA-6 & 0h 22m 24.83s & +0d 20m 11.8s  & 1.8 & 22.63 & -0.72  & 1.04  & 1.06 & 1.33 & 3.71 & 0.98  &1.32   \\
WHDF-EVLA-7 & 0h 22m 39.94s & +0d 18m 38.6s  & 0.6 & 22.44 & 1.20  &  3.42  & 1.67 & 2.05 & 4.27 & 0.90  & -  \\
WHDF-EVLA-8 & 0h 22m 22.81s & +0d 20m 14.6s  & 0.9 & 24.02 & -1.25  &  0.79  & 0.71 & 1.12 & 3.44 & 0.85  & 2.11  \\
WHDF-EVLA-9 & 0h 22m 32.56s & +0d 23m 27.6s  & 1.1 & 19.31 & -0.29 & 1.62 & 0.78 & 1.04 & 2.72 & 0.76  & 0.382  \\
WHDF-EVLA-10 & 0h 22m 20.48s & +0d 19m 0.3s  & 1.4 & 20.15 & -0.18 &  1.76 & 0.86 & 1.04 & 2.82 & 0.78  & 0.432  \\
WHDF-EVLA-12 &0h 22m 11.08s&+0d 19m 50.9s& 2.0&24.18& -0.33 &  0.48& 0.47 & - &3.85 & - &  -\\
WHDF-EVLA-14 & 0h 22m 24.99s&+0d 16m 47.8s&2.5&21.74& -0.26 & 1.58 & 0.43 & 0.57 & 1.56& - &  -\\
WHDF-EVLA-16 & 0h 22m 14.86s & +0d 21m 4.2s  & 1.1 & 20.50 & 0.06 &  1.92 & 0.88 & 2.06 & 3.16 & -  & -  \\
WHDF-EVLA-17 & 0h 22m 42.84s & +0d 19m 54.8s  & 2.3 & 21.60 & 0.02 &  2.24 & 0.96 & 1.16 & 2.92 & 0.75  & -  \\
WHDF-EVLA-18 & 0h 22m 37.79s & +0d 23m 38.1s  & 1.3 & 17.39 & -0.14 & 1.31 & 0.70 & 0.82 & 2.42 & 0.56  & -  \\
WHDF-EVLA-19 & 0h 22m 22.62s & +0d 20m 43.0s  & 0.6 & 22.12 & -0.27 & 0.87 & 0.56 & 0.89 & 2.72 & 1.11  & -  \\
WHDF-EVLA-20 & 0h 22m 53.22s &+0d 16m 59.8s & 4.3 & 18.43 & 0.06 & 1.69 & 0.78 & 0.80 & 2.57 & - &  -\\
WHDF-EVLA-23 & 0h 22m 32.09s & +0d 18m 11.1s  & 1.3 & 24.91 & 0.16 & 1.29 & 1.36 & - & 4.79 & 0.73 & -  \\
WHDF-EVLA-25$^*$ & 0h 22m 36.78s & +0d 20m 56.5s & 1.5 & 19.82 & - & - & - & - & - &1.01 &  -\\ 
WHDF-EVLA-27 & 0h 22m 16.22s & +0d 19m 45.7s & 1.4 & 21.39 & -0.21 & 2.01 & 0.78 & 1.06 & 2.51 & - & - \\
WHDF-EVLA-29 & 0h 22m 21.80s & +0d 21m 47.3s & 3.2 & 22.29 & -0.68 & 0.83 & 0.91 & 0.66 & 2.52 & 0.88 & - \\
WHDF-EVLA-32 & 0h 22m 28.24s & +0d 21m 53.5s & 5.0 & 22.62 & -0.17 & 0.97 & 0.41 & -0.11 & 1.51 & - & - \\
WHDF-EVLA-39 & 0h 22m 30.95s & +0d 22m 54.5s  & 1.8 & 23.77 & -0.41 & 1.47 & 1.00 & 1.31 & 3.63 & 1.26 & -  \\
\hline
\multicolumn{12}{|l|}{$^*$ Position and magnitude are taken from the H-band image.}
\end{tabular}
\caption{Optical and near-infrared properties of the nearest counterpart
within $5''$ of the positions of the radio sources in Table
\ref{tab:sources}. Magnitude, colours and positions are taken from
imaging described in Metcalfe et al.~(2001) and Metcalfe et al.~(2006).
Spectra are unpublished data taken with LDSS2 on the Magellan 6.5m
telescope.}\label{tab:optical}
\end{minipage}
\end{table*}

\begin{table*}
\begin{minipage}{170mm}
\centering
\begin{tabular}{lclcccccc} \hline
Radio ID& LABOCA        & Comment   & R  & U-B& 870$\mu$m  & 8.4GHz & LAB-EVLA & Redshift  \\
        &     ID       &           & mag&    & (mJy)     &  ($\mu$Jy)& Separation  &  \\ 
\hline

WHDF-EVLA-6  &LAB-11& absorbed QSO/UVX & 22.63 & -0.72  & 3.4 & 48 & $7.''7$ & 1.33\\
WHDF-EVLA-8  &LAB-05& absorbed QSO/UVX & 24.02  & -1.25  & 4.0 & 37 &$3.''1$ & 2.12\\
WHDF-EVLA-15 & -    & blank            & -     &    -    & -   & 28 & -& -  \\
WHDF-EVLA-19 & -    & merger-train     & 22.12 & -0.27  & - & 23 & -& -  \\
WHDF-EVLA-23 & -    & stellar - QSO?   & 24.91 &  0.16  & - & 47 & -&- \\
WHDF-EVLA-25$^*$& - & i dropout        & 19.82 &    -   & - & 32 & -& - \\ 
WHDF-EVLA-27 & -    & merger-double    & 21.39 & -0.21  & - & 39 & -& -\\
WHDF-EVLA-29 & -    & blank            & -     &    -    & -   & 19 & -& -  \\
WHDF-EVLA-32 &LAB-02& blank            &   -   &    -   & 4.3 & 15 & $5.''4$&-  \\
WHDF-EVLA-39 & -    & merger-train/UVX & 23.77 & -0.41  &-& 18 &- &-  \\
\hline
WHDF-EVLA-1  &LAB-06& spiral galaxy   & 16.07 & 0.04  & 3.9 & 201 & $5.''5$ & 0.046\\
\hline
\multicolumn{5}{|l|}{$^*$ Magnitude is taken from the H-band image.}
\end{tabular}
\caption{Summary of Tables \ref{tab:sources}, \ref{tab:optical}
and Table 1 of \citet{bielby2012} for complete sample of 10 faint EVLA
sources with $S<50\mu$Jy at 8.4GHz. For completeness of the sub-mm
fluxes and LABOCA-EVLA separations, WHDF-EVLA-1 is also listed, although
too bright at 8.4GHz for inclusion in the faint sample.
}\label{tab:faint}
\end{minipage}
\end{table*}

\begin{figure*}
\vspace{180mm}
\centering
\setlength{\unitlength}{1cm}
\subfloat{\includegraphics[width= 0.8 \textwidth]{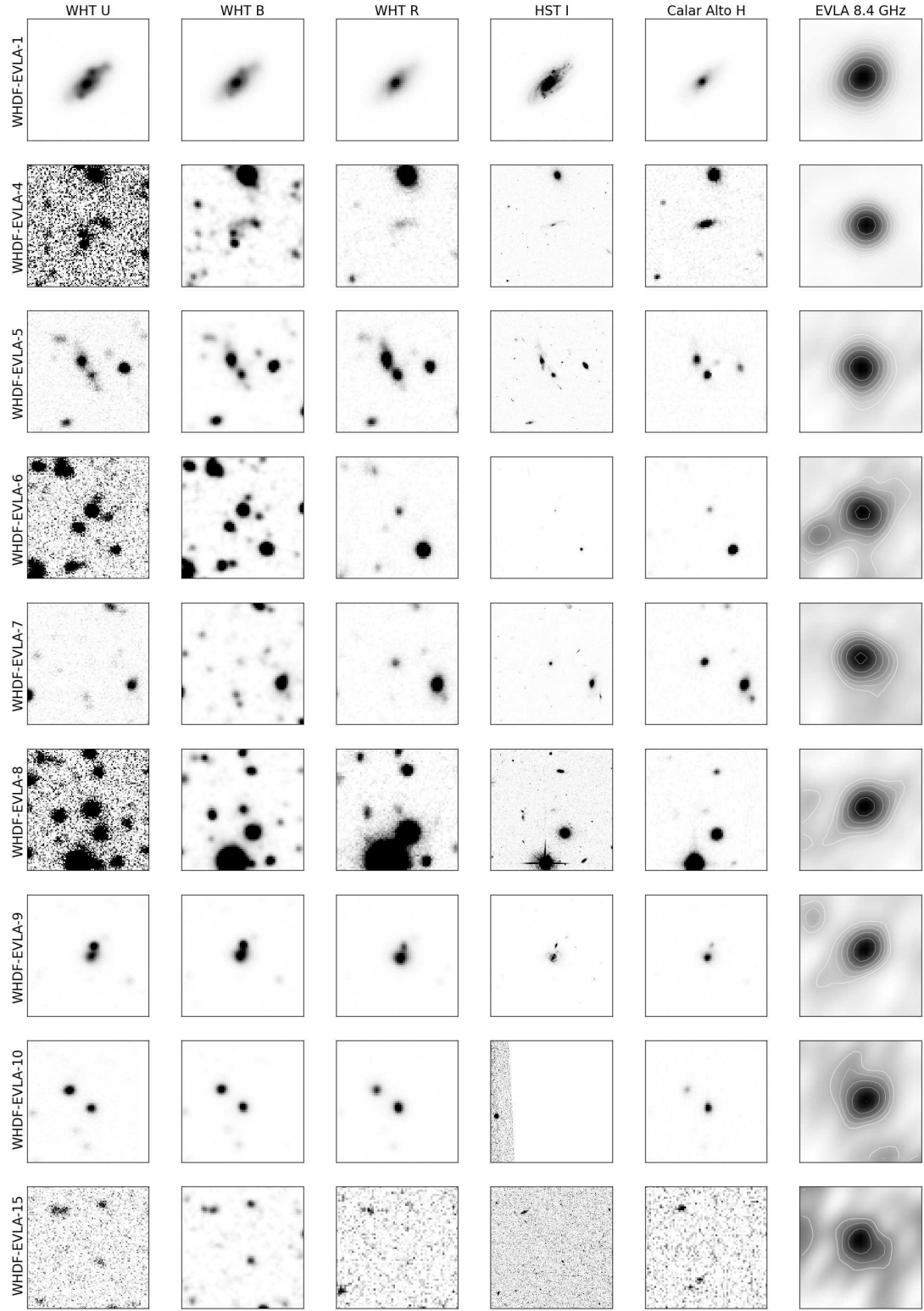}}
% \begin{picture}(15,22.5)
% \put(-1.1,-0.6){\special{psfile=evla_optical_0.eps vscale=82.0 hscale=82.0 voffset=0
% hoffset=0}}
% \end{picture}
\caption{\label{fig:evla_optical} Near infrared / ultraviolet thumbnail
images centred on the EVLA radio sources within the central area of the
WHDF. The radio sources are presented in the right hand column.
Photometric wavebands and the instrument used are noted above each
column, radio source IDs are noted at the end of each row. The base
contour level on the radio cutouts is 2$\sigma$ and increases in
multiples of $\sqrt{2}$. Each thumbnail spans 25 arcseconds. This figure continues on the next page.}
\end{figure*}

\begin{figure*}
\vspace{120mm}
\centering
\setlength{\unitlength}{1cm}
\includegraphics[width= 0.8 \textwidth]{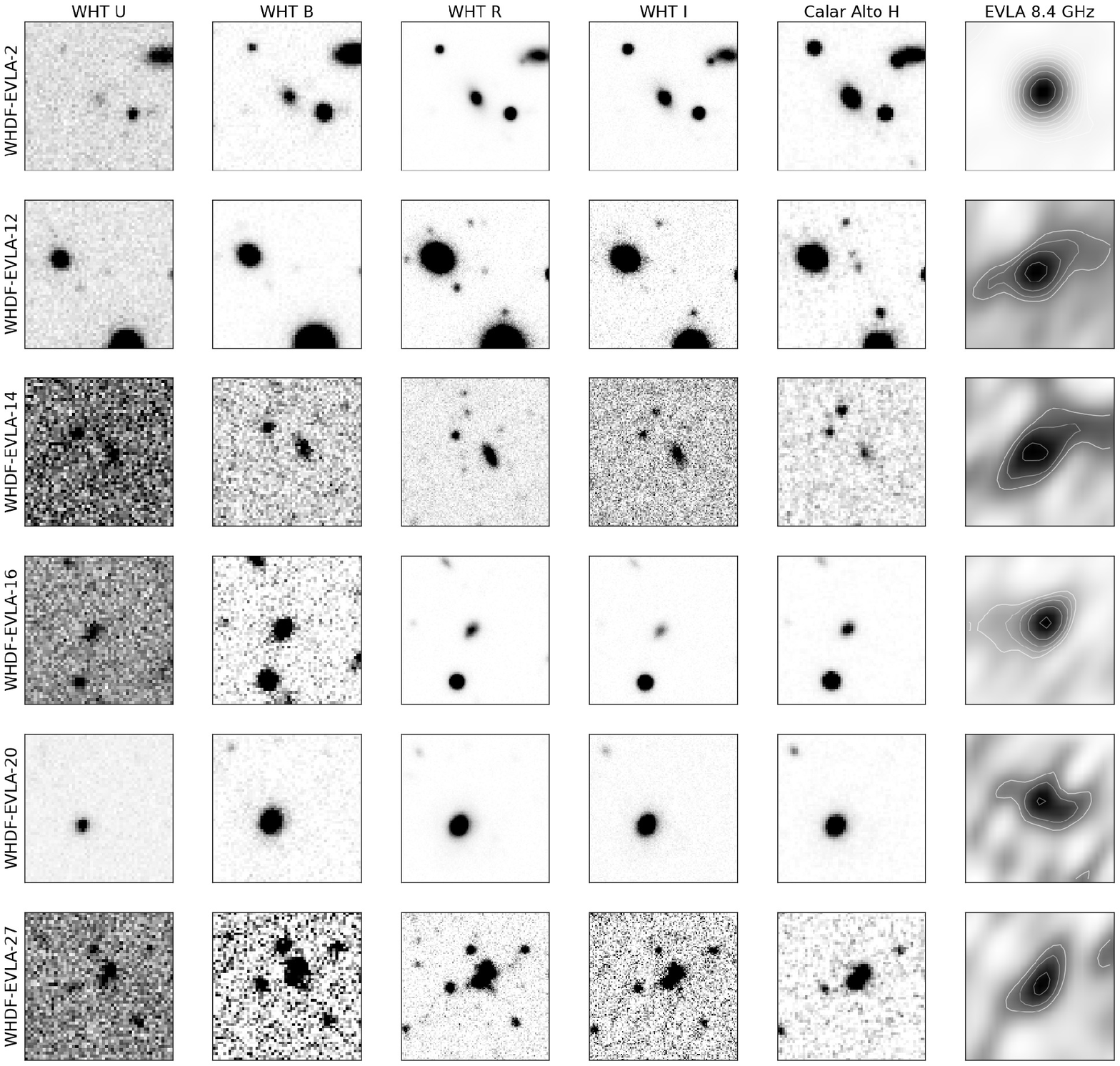}
% \begin{picture}(15,22.5)
% \put(-1.1,-0.6){\special{psfile=evla_optical_0.eps vscale=82.0 hscale=82.0 voffset=0
% hoffset=0}}
% \end{picture}
\caption{\label{fig:evla_optical_ext} Near infrared / ultraviolet
thumbnail images centred on the EVLA radio sources within the extended
area of the WHDF. As per Figure \ref{fig:evla_optical}, photometric
wavebands and the instrument used are noted above each column. Radio
source IDs are noted at the end of each row and the radio sources
themselves are shown in the right hand column. The base contour level on
the radio cutouts is 2$\sigma$ and increases in multiples of $\sqrt{2}$.
Each thumbnail spans 25 arcseconds.}
\end{figure*}

This section presents possible (near-)infrared, visible and ultraviolet
counterparts for the radio sources and discusses noteworthy individual
sources. Cut-out images from existing observations are plotted for each
radio source in the central 7'~$\times$~7' area of the WHDF in Figure
\ref{fig:evla_optical}, and in the extended area in Figure
\ref{fig:evla_optical_ext}. Photometric bands U (365~nm), B (445~nm), R
(658~nm), I (806~nm) and H (1630~nm) are shown, with the telescope used
indicated above each column. The final column shows the radio image with
overlaid contours. The base contour level is 2$\sigma$ and increases in
multiples of $\sqrt{2}$. Each thumbnail spans 25 arcseconds. The nearest
counterparts to the radio sources are listed in Table \ref{tab:optical}.
Photometric colours are listed, as are redshifts, where available.

\noindent {\bf WHDF-EVLA-1:} The brightest radio source (aside from 4C
00+02) in the central region of the WHDF corresponds to a spiral galaxy at $z$~=~0.046
which is also associated with a sub-mm LABOCA detection and an X-ray
source.

\noindent {\bf WHDF-EVLA-3:} This source is a low redshift (z~=~0.2609) galaxy.

\noindent {\bf WHDF-EVLA-4:} Near infrared imaging of this source shows
an edge-on spiral which is clumpy in ultraviolet imaging.

\noindent {\bf WHDF-EVLA-6:} This radio source corresponds to one of the
LABOCA-detected obscured quasars at $z$~=~1.33 and a faint infrared
counterpart can be seen at the centre of the major radio peak. The radio
emission is extended and resolved into two components. There is no clear
optical or infrared counterpart for the lower radio peak suggesting that
the quasar may host a radio jet. Paradoxically, the optical colours 
show an ultraviolet excess (UVX) while the R-H colours are quite red. At HST
resolution the source may just be resolved. The redder near-infrared colours may 
be explained by host domination in these bands. Assuming all the R band flux comes 
from the QSO and R-K~$\approx$~2 (e.g.~Maddox et al., 2012), the QSO would have K~$\approx$~20.6 implying K~$\approx$~18.1 for the host galaxy. Such galaxy magnitudes are not uncommon at this redshift in surveys such as the K20 survey (Cimatti et al., 2002), and as evidenced further by the
K-band luminosity function derived from the UKIDSS Ultra Deep Survey (Cirasuolo et al., 2010).
Note also the general tendency for AGN at all epochs to lie in the most massive galaxies (Dunlop et al., 2003).

\noindent {\bf WHDF-EVLA-8:} The second LABOCA-detected obscured quasar
at $z$~=~2.12. In the HST I band image the object has no nucleus and
looks like an edge-on spiral, but a nucleus is seen both in the blue and
the near-infrared imaging. Again, the colours of the quasar are anomalous, having
U-B = -1.25, i.e.~very blue, whereas R-H = 3.44 i.e.~very red. Again, the
redder near-infrared colours may be explained by increasing host domination in these bands.
On the same assumptions as for WHDF-EVLA-6 the host galaxy could have K~$\approx$~19.8
and galaxies are seen out to z~$\approx$~2 at this limit in the K20 survey.

\noindent {\bf WHDF-EVLA-9:} This is a low-redshift spiral galaxy
($z$~=~0.382) which harbours an optical jet and has associated
X-ray emission.

\noindent {\bf WHDF-EVLA-19:} This object shows a stellar nucleus in the ultraviolet, is nearly
UVX and appears among a string of fainter objects. It is a possible quasar or merger.

\noindent {\bf WHDF-EVLA-23:} This source has a stellar appearance in the HST I band images. It is very red
at R-H and H-K. The U-B colour is blue but not UVX so it could be a $z\ga2.2$ QSO
or a modestly reddened QSO at lower redshift, perhaps contaminated in
the infrared by the host galaxy.

\noindent {\bf WHDF-EVLA-25:} This is a drop-out at I-band and other optical bands. 
It is only detected in H imaging, where it is relatively bright.

\noindent {\bf WHDF-EVLA-27:} This is a possible double object, a candidate merger with relatively blue colours.

\noindent {\bf WHDF-EVLA-39:} The UVX here implies probable quasar. Infrared and HST I-band imaging
resolves this source into multiple components so it could be a merger. Otherwise, a possible strong
gravitational lens? In the latter case, the uppermost pair of images may form a partial Einstein ring.

\begin{figure*}
\vspace{154mm}
\centering
\setlength{\unitlength}{1cm}
\subfloat{\includegraphics[width= 0.8 \textwidth]{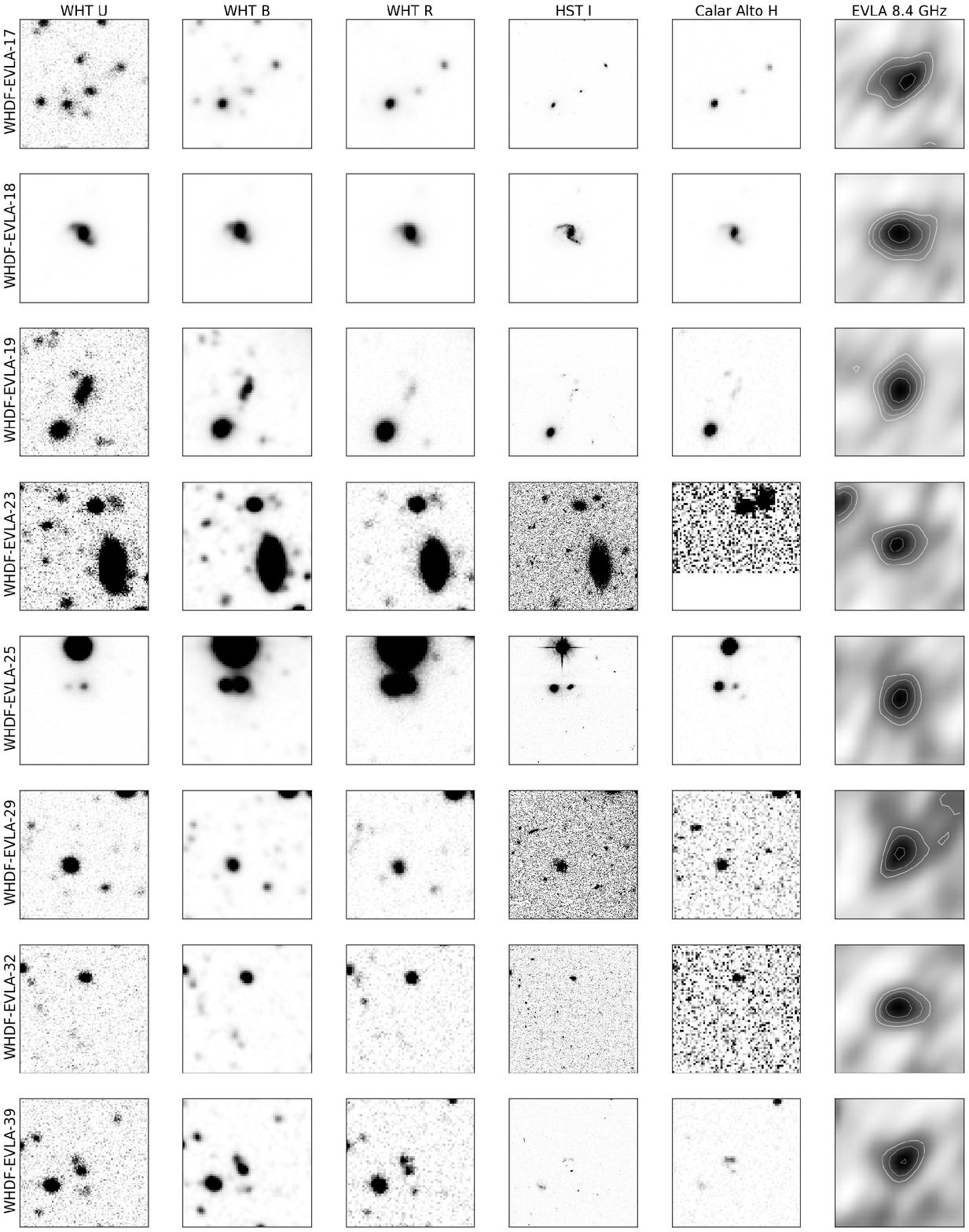}}
% \begin{picture}(15,22.5)
% \put(-1.1,-0.6){\special{psfile=evla_optical_0.eps vscale=82.0 hscale=82.0 voffset=0
% hoffset=0}}
% \end{picture}
\end{figure*}

Of the 12 sources within the $\approx5'$ half-power point, 9 have
optical/near-infrared counterparts within $2''$ and 3 are UVX. 4 are also LABOCA
sources. Similarly, in the full area, of the complete sample of  10
faint ($<50\mu$Jy) sources (see Table \ref{tab:faint}) 6 have
optical/near-infrared counterparts of which 3 are UVX (i.e.~likely quasars) including
the two absorbed quasar LABOCA sources. The other 3 faint radio
counterparts are not UVX but are only slightly less blue and  likely to
be star-forming galaxies, predominantly at lower luminosities and
redshifts. Of the 10, there are 3 possible mergers, one of which
contains a UVX source. The 4 faint optically unidentified radio sources
may be either dust obscured quasars or galaxies. One of these sources is
identified only at H and K and it appears resolved even at ground-based
resolution.

The LABOCA 870$\mu$m fluxes of the four radio sources with LABOCA counterparts 
are also given for completeness in Table \ref{tab:faint}. Their high
ratio of sub-mm relative to radio flux suggests that all four sources are likely to be 
dust rather than synchrotron dominated in the sub-mm.

\subsection{Faint radio source counts at 8.4~GHz}

Determining the counts of extragalactic sources at radio wavelengths has
been an active area of study for several decades; de Zotti et al.~(2010)
present a review of both the history and the state of the art. Early
radio surveys provided a key fulcrum in Steady State versus Big Bang
cosmology debates in the 1950s, and observations since then have
revealed much about cosmology and the evolution of radio sources with
cosmic time, and have (particularly at higher frequencies) proved
essential for categorising extragalactic foregrounds for Cosmic
Microwave Background experiments. Source counts at the faint
($\leq$1~mJy) end of the distribution exhibit a turn-up. This is
generally explained by the increase in the dominance of star-forming
galaxies over AGN at these low luminosities (e.g.~Padovani et al., 2009)
and it has also been claimed that radio-weak AGN make a significant
contribution (e.g.~Jarvis \& Rawlings, 2004, Smol{\v c}i{\'c} et al., 2009, Simpson 
et al., 2006, 2012), although the exact nature of
this excess remains a source of debate. While most studies of source
counts derived from radio surveys have been conducted at L-band, the
turn up persists in higher frequency observations including those such
as the X-band observations presented in this paper.

The recent Absolute Radiometer for Cosmology, Astrophysics and Diffuse
Emission (ARCADE2) experiment (Fixsen et al., 2011) has also piqued
interest in the faint end of radio source populations due to the
measured excess in the sky brightness temperature at 3 GHz (Seiffert et
al., 2011). Vernstrom et al.~(2011) use source count data from 150 MHz
to 8.4 GHz to predict the contribution to the sky temperature background
from measured source populations and conclude that if the ARCADE2 result
is correct there must be an additional significant population of radio
galaxies at fluxes fainter than those hitherto reached by radio surveys.
This clearly motivates the need for deeper radio continuum surveys.

The source counts derived from the EVLA observations of the WHDF do not
reach the depths needed to begin to address the ARCADE2 result; the counts are presented here for
completeness. It is noteworthy however that the depth of the
observations in this paper (2.5 $\mu$Jy) is approaching that of the
deepest 8.4 GHz observations (1.49~$\mu$Jy in the SA13 field; Fomalont
et al.~2002) which required 190 hours of integration time with the old
VLA system. The survey speed advantage is brought about by the huge
increase in available bandwidth ($\Delta \nu$), since the noise level in
a synthesis image $\sigma \propto \Delta \nu^{-0.5}$. Increasing the
bandwidth of the observation does not change the traditional limiting
factor for high frequency survey work, which is that the sky area
covered by a single pointing is proportional to $\nu^{-2}$ where $\nu$
is the observing frequency, but it does mean that far less time is
required to reach a certain depth per pointing. High frequency surveys
covering significant sky areas thus become far more economical. Note
that since the WHDF radio observations were taken the bandwidth
available for general users of the EVLA has increased by a factor of 8.

\begin{table}
\centering
\caption{Source counts and bin details for the data plotted in Figure \ref{fig:source_counts}.\label{tab:source_counts}}
\begin{tabular}{llll} \hline
Bin & Central flux ($\mu$Jy) & Bin width ($\mu$Jy) & $N$ \\ \hline
1 & 15.2  & 10.4  & 3\\ 
2 & 30.9  & 22.0  & 5\\
3 & 62.9  & 43.0  & 4\\
4 & 128.2& 87.5  & 3\\
5 & 260.9& 178.1 & 2\\ \hline
\end{tabular}
\end{table}

\begin{figure}
\vspace{83mm}
\nonumber
\centering
\includegraphics[width= \columnwidth]{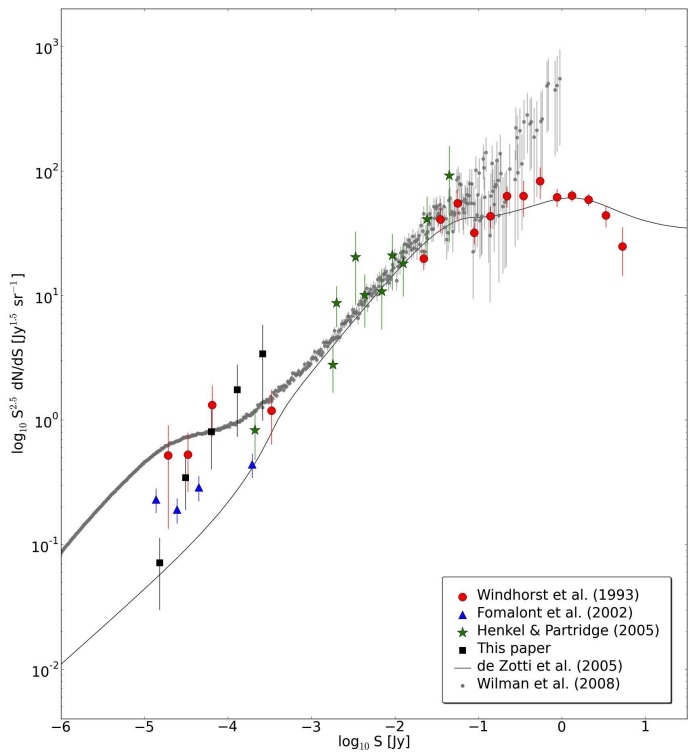} 
\caption{\label{fig:source_counts}Euclidean-normalized differential
source counts at 8.4 GHz. Observational data are from Windhorst et al.~(1993),
Fomalont et al.~(2002) and Henkel \& Partridge (2005), collated by de
Zotti et al.~(2010), and also from this paper. Only sources which have
been corrected for primary beam attenuation were used when determining
the source counts for the latter. Sources were placed into five bins of
logarithmically increasing width between 10 and 350 $\mu$Jy, the lower
limit being equivalent to the completeness limit. The error bars on the
counts from this paper are derived from Poisson statistics. Simulated source counts are from
de Zotti et al.~(2005) and from Wilman et al.~(2008).}
\end{figure}

The source counts from the WHDF observation are shown in Figure
\ref{fig:source_counts} along with three other previously published data
sets. The counts are normalized to those expected in a
non-expanding Euclidean universe for ease of comparison to the published
data. The origins of the previously published points are shown on the
figure, the values of which were collated and listed by de Zotti et al.
(2010). Only the 17 sources from the WHDF observations which could be
reliably corrected for beam attenuation effects were included in the
count. Omitting this step would clearly artificially bias the counts
towards fainter levels. Sources were counted in five bins between 10 and
350~$\mu$Jy with logarithmically increasing widths. Error bars are
simply derived from Poisson statistics. The solid angle covered by the
single pointing within the region where the beam gain is greater than a
factor of 0.2 is 0.0119 deg$^{2}$ for the central frequency.

For reference, also shown on Figure \ref{fig:source_counts} are the
source counts predicted by two models down to a flux limit of 1~$\mu$Jy.
The small circles show the source counts generated by binning all
$\sim$260 million galaxies in the full 20~$\times$~20 deg$^{2}$ sky area
of the semi-empirical galaxy simulation of Wilman et al.~(2008). Linear
extrapolation of the 4.86 and 18~GHz fluxes offered by the simulation
was used to generate the source count data. The solid line shows the
model of de Zotti et al.~(2005), solely for AGN-powered sources. The
faint count data lie above the AGN model, illustrating the need for
another component at faint fluxes. In the Wilman et al. (2008) model, this is
represented by star-forming galaxies (67\%) and also radio-weak AGN (20\%),
with the latter therefore making a comparable contribution to radio-loud AGN (13\%)\footnote{These fractions were
determined by extracting a sample of galaxies with fluxes between 10 and 50 $\mu$Jy
from 25 square degrees of the Wilman et al.~(2008) simulation. These 100444 sources
were then binned according to their star formation and AGN types: 20508 radio quiet AGN; 11694 FR-I; 0 FR-II; 1069 gigahertz peaked spectrum; 55894 quiescent star forming galaxies; 11279 starburst galaxies.}.
The apparent overestimation of the total model counts at the faintest limits may
explain the more $\approx$1:1 split of AGN (both radio-loud and
-weak) and star-forming/merging galaxies seen in Table 2 compared to the $\approx$2:1 ratio
in favour of star-forming galaxies predicted by the model.
Further discussion of the nature of the faint radio
source population in the WHDF is presented in Sections
\ref{sec:discussion} and \ref{sec:conclusions}.

\section{Discussion}
\label{sec:discussion}

The EVLA survey of the WHDF confirms the associations between
LABOCA sub-mm sources and X-ray absorbed AGN.  Two of the three X-ray
sources associated with LABOCA sources  are hard X-ray sources and
likely to be cold gas absorbed, with the third X-ray source being
associated with the nucleus of a nearby galaxy and too faint to measure
its hardness ratio. All three sub-mm$+$X-ray sources are associated with
radio sources. Although the numbers are small, the X-ray fraction is
high and so is the gas-absorbed fraction. Previously,  Page et
al.~(2004), who studied samples of absorbed (density of Hydrogen atoms
$N_{H}$~$>$~10$^{22}$~cm$^{-3}$) and unabsorbed quasars, suggested a
similar difference in the sub-mm properties of these two classes of
sources. \citet{hill2011a} themselves found that the LABOCA ECDFS sub-mm
Survey (LESS, \citealt{wardlow2011}) showed clear evidence for
$\approx20$\% of sub-mm sources being associated with X-ray absorbed AGN.

Models whereby absorbed AGN are used to fit the X-ray background with the 
absorbed photons being re-radiated in the infrared have been
discusssed by \citet{gunnphd} and \citet{gunn1999} and more recently by
\citet{hill2011a}.  In these models, as the absorbed  X-ray fraction
increases with hydrogen column density, the re-radiated far-infrared flux increases in
proportion and they therefore predict that there will be more sub-mm
radiation from X-ray absorbed AGN. Such explanations therefore have
implications for the unified model (e.g.~Antonucci, 1993) which holds that the differing X-ray
and optical properties of obscured and unobscured AGN can be explained
in terms of the inclination of the system with respect to the line of
sight of the observer. The  amount of sub-mm emission, originating in the dusty
torus surrounding the system, should then be independent of  its orientation. The
differing sub-mm properties between two classes of AGN thus cannot be
explained by invoking the unified model. Previously it was suggested
that sub-mm-bright AGN are being observed during a different
evolutionary phase, whereby the AGN experiences a period of growth
within a dusty, starforming galaxy environment (Page et al., 2004). It is speculated here 
that X-ray absorbed AGN might be expected to
be strong sub-mm sources if the presence of cold, neutral gas in an AGN
powered source also indicated the presence of cold dust. In this case 
the dust torus would need to extend up to $\sim$1~kpc to maintain a  dust 
temperature as low as 30--35~K (e.g. \citealt{kura2003}).

It is also noted that the optical and near-infrared colours of the two X-ray absorbed
quasars are anomalous, being quite UVX in U-B yet quite red in R-H. If this
is not due to variability then it might suggest that the optical colours
are active nucleus dominated whereas the near-infrared colours are host dominated.
Clearly this could favour a picture where the nuclear sight line is
cold gas absorbed as evidenced by both the hard X-rays and the optical
narrow lines. But the dust would then have to form a more clumpy or 
toroidal structure around the nucleus to leave the nuclear sight-line  unobscured by dust.
The fact that narrow, high-ionisation lines like C~IV are seen suggest the 
gas absorption is on smaller scales than the narrow line region, however
there appears to be little dust on the nuclear sightline. This behaviour
may be more characteristic of a unified model but such a model does not 
explain the basic sub-mm - neutral gas absorption correlation. A 
non-unified model which does explain this correlation would then also
need to invoke a mechanism, maybe a jet, to destroy cold dust  just
along the nuclear sightline while leaving the neutral gas more 
or less in place. It is interesting that the sub-mm absorbed X-ray AGN, WHDF-EVLA-6, does
show evidence of a radio jet.

The two absorbed QSOs also have radio and sub-mm  fluxes that put them close to
the FIR-radio correlation (e.g. Jarvis et al. 2010 and references therein). For example,
WHDF-EVLA-08 at $z$~=~2.12 has S$_{870\mu m}$~=~ 6.9~$\times$~10$^{25}$~WHz$^{-1}$
and L$_{8.4GHz}$~=~8.6~$\times$~10$^{23}$~WHz$^{-1}$. An approximate conversion
assuming dust temperature components of 35-60~K in the wavelength
range 8-1000~$\mu m$ gives L$_{FIR}$~$\approx$~1~$\times$~10$^{39}$~W~=~ 
3~$\times$~10$^{12}$~L$_{\odot}$. Similarly, an approximate conversion of the
radio flux assuming a $\nu^{-1}$ spectrum gives
L$_{1.4GHz}$~$\approx$~5~$\times$~10$^{24}$~WHz$^{-1}$. This gives
q$_{IR}$~$\approx$~1.75 (see Jarvis et al 2010) compared to the average
for star-forming galaxies of q$_{IR}$~$\approx$~2.2, indicating some degree
of contamination  by non-thermal radio emission.

\citet{hill2011a} have argued that in a non-unified model the absorbed AGN
may make up a significant fraction of the bright  sub-mm sources.
Also, Mart{\'{\i}}nez-Sansigre et al.~(2005) have inferred from radio
observations that most black hole accretion is obscured, implying
that the contribution of obscured AGN to the sub-mm background may be
very significant.

This non-unified picture where absorbed AGN are preferentially sub-mm
loud is also supported by several more recent results. For example,
\citet{page2012} find that most of the AGN detected in the Chandra Deep
Field North at 250~$\mu$m by \emph{Herschel} SPIRE show strong X-ray
absorption, although these authors emphasised more the lack of
far-infrared emission from the intrinsically bright X-ray AGN.
\citet{simpson2012} have also reported that two high redshift 2SLAQ
quasars detected in \emph{Herschel} ATLAS as strong far-infrared sources
appear to show heavily absorbed Lyman-$\alpha$ lines.  Rovilos et
al.~(2012, priv. comm.) using Herschel PACS/SPIRE have also found that
the cold dust component in ECDFS AGN is as significant a component in
AGN spectral energy distributions (SEDs) as their hot dust
components. \citet{hickox2012} have also found that the galaxy group
environment and clustering of sub-mm sources are completely  consistent
with those of QSOs.  Finally, a criticism
of the \citet{hill2011a} model is that the dust mass required will have
to be around 5~$\times$~10$^{6}$~M$_\odot$ for L$_{*}$ QSOs, implying a cold gas
mass of perhaps $\approx$5~$\times$~10$^{8}$~M$_{\odot}$. It is interesting to note
that \citet{molinari2011} have detected a $\approx$100~pc ring with
$\approx$5~$\times$~10$^{7}$~M$_{\odot}$ of gas and dust around the Sgr A black hole
candidate in the centre of the Milky Way. If this was associated with a
$\approx$10$^{43}$~ergs$^{-1}$ AGN outburst then it would at least be of
the right size and mass to produce a cold sub-mm component in the Milky
Way.

A more typical non-unified model invokes an evolutionary relation
between absorbed and unabsorbed QSOs (e.g.~Sanders et al., 1988, Fabian
1999). Here a merger triggers the birth of an obscured QSO, accompanied
by a burst of star-formation. Once the spheroid is formed, expulsion of
cool gas and dust quenches the star-formation, turns off the sub-mm
emission and leaves the QSO unobscured. This model implies that more
absorbed QSOs will be more FIR/sub-mm bright. Such a model would thus
also be consistent with X-ray absorbed QSOs being preferentially sub-mm
bright as observed in this paper. The main difference with the model of
\citet{hill2011a} is that there the dust is AGN heated whereas in this
other case it is starburst heated. \citet{hill2011a} argued that 
starburst heating leaves the similarity of the sub-mm sources'
luminosities to that of QSOs looking like a coincidence. High resolution
ALMA observations could also measure the size of the sub-mm emitting
region and help differentiate between these possibilities.

Herschel ATLAS results from Bonfield et al.~(2011) find a
correlation between optical QSO and FIR luminosity at fixed redshift as
well as with redshift. This is consistent with the model of Hill \&
Shanks (2011a) since at fixed column and redshift, higher optical/X-ray
luminosity implies higher FIR luminosity. The amount of sub-mm emission
from QSOs with relatively low absorption of log(N$_H$)~$\approx$~20.5 or
A$_V$~$\approx$~0.1-0.2~mag, implies that 10-20\% of the QSO optical luminosity
will be re-radiated by dust, implying no disagreement with the substantial 
FIR luminosities for these objects as observed by Bonfield et al.~(2011; see also
Hatziminaoglou et al.,~2010, Serjeant et al.,~2010).

For QSO dust components at higher temperatures measured  at
shorter wavelengths, there may be  more evidence for a unified model from
finding correlations between AGN radio and IR/FIR dust emission if  the radio
flux is assumed to be orientation independent. Shi et al.~(2005) found
some indication of a correlation between  $70\mu$m and radio  fluxes for
a sample of AGN observed with Spitzer MIPS. Previously, Polletta et 
al.~(2000) found that in a sample of 22 AGN,  the amount of IR/FIR  
dust emission was reasonably  independent of whether the AGN were radio loud or not, 
again broadly consistent with a unified picture. Whether these results 
present a problem for either of the non-unified scenarios discussed above 
awaits further data.

The identification of optical/near-infrared counterparts for 
10 faint ($<$50~$\mu$Jy) radio sources show a mixture of $\approx30$\%
lower redshift, blue, starforming/merging radio galaxies (3/10) and
$\approx30$\% higher redshift quasars/AGN (3/10), including the absorbed
quasars identified as sub-mm sources. The remaining 4/10 faint sources are
optically blank, although one is detected in the near-infrared as a resolved
source. Otherwise, there are few sources identified with red, early-type
galaxies. Thus it might be concluded that this mix of sources at 8.4 GHz
is not dissimilar to the mix of sources found in the sub-mm samples at
350~GHz. Indeed, \citet{hill2011b} suggested that this mix persists to
the \emph{Herschel} SPIRE bands at 350~$\mu$m, i.e.~850~GHz, with the AGN
component then becoming increasingly less dominant at higher
frequencies. The results presented above seem to support the idea that a strong
AGN component is seen at all frequencies from the radio to the far-infrared, with
the AGN component synchrotron-dominated at low frequencies and
increasingly cold dust-dominated at higher frequencies.

\section{Conclusions}
\label{sec:conclusions}

Radio continuum observations can now achieve extreme depths with
relatively short integration times due to the vastly increased
correlator bandwidths that are now available. This makes surveys much
more economical, thus prospects are good for future, deep high-frequency
surveys covering significant sky areas with instruments such as the EVLA
and MeerKAT (e.g.~Jarvis, 2011; Heywood et al., 2011). There is a caveat here however: both
deeper sensitivity limits and broad bandwidths conspire to increase the
calibration complexity. Direction dependent effects (which are often
also highly frequency-dependent) need to be properly accounted for
during calibration in order to remove the artefacts associated with
off-axis sources. As observations are pushed deeper these effects become
more apparent (see also \citealt{smirnov2011c}), and if the science targets are
extremely faint they are in danger of being swamped by residual
calibration errors. It is fair to describe the location of the confusing
source which blighted the WHDF observations as the `worst-case scenario'
(see Appendix for details) yet the success with which it was removed is
encouraging if future deep radio continuum surveys are to routinely
produce images which are limited by thermal noise as opposed to
artefacts brought about by deficiencies in either the model of the sky
or the instrument.\footnote{``If it's bright enough to cause trouble
it's bright enough to be solved for."-- J.~E.~Noordam, \emph{ipse
dixit}.}

Applying these techniques to EVLA data, a deep 8.4 GHz radio image of an
area covering the Extended WHDF has been generated, and a catalogue of
41 radio sources with flux densities exceeding 4$\sigma$ has been
derived. The central, deepest area of the WHDF is well matched to the
main lobe of the EVLA primary beam, and within this area 17 sources are
detected that can have their apparent fluxes corrected for the beam
attenuation.

Two of these radio sources (WHDF-EVLA-6 and WHDF-EVLA-8) confirm the
association of two high redshift ($z$ =~1.33 and 2.12 respectively)
X-ray absorbed quasars with the sub-mm sources detected by
\citet{bielby2012}. Such sources  warrant further investigation due to
the uncertainty surrounding the contributions that AGN make to the
sub-mm background. Both the sub-mm absorbed quasars show unabsorbed nuclear
colours in the blue that would require a mechanism to remove cold dust
but not neutral gas from the quasar sightline. It is interesting that one
of the sub-mm sources, WHDF-EVLA-6, appears to harbour a radio jet.
Certainly the EVLA $+$ LABOCA  results from  the WHDF support the
previous result of \citet{hill2011a}  from ECDFS/LESS that X-ray
absorbed AGN are significant sub-mm sources. This in turn may support
their non-unified absorbed AGN X-ray background  model where AGN dust
emission is proportional to their X-ray column which predicts that up to
$\approx$40\% of the sub-mm background may be due to AGN.

The beam-corrected source fluxes are used to determine differential
source counts which are in good agreement with previously published
values covering the distribution down to $\approx$50-100~$\mu$Jy and the
semi-empirical extragalactic simulation of Wilman et al.~(2008). The
counts show a significant excess over the purely AGN-powered count model
of de Zotti et al.~(2005). The optical identifications suggest the WHDF
faint source counts are composed of 30\% high redshift AGN, 30\% low
redshift star-forming galaxies and 40/30\% optical/near-infrared blank
fields, likely either to be dusty star-forming galaxies or the more
heavily absorbed AGN needed to explain the X-ray background. Various
individually interesting counterparts within this faint radio source
population have been noted.

Future EVLA observations over a wider area of the WHDF will test how
representative the results for the present area are. Deeper optical and
near-infrared observations are needed to determine further the nature of the
population of faint radio sources responsible for both the turn-up in
faint source counts and the excess radio sky temperature detected by the
ARCADE2 experiment.

\section*{Acknowledgments}

We thank the anonymous referee for very useful comments. The National Radio Astronomy Observatory is a facility of the National Science Foundation operated under cooperative agreement by Associated Universities, Inc. IH thanks The South East Physics Network (SEPnet). This work is based upon research supported by the South African Research Chairs Initiative of the Department of Science and Technology and National Research Foundation. We thank Gianfranco de Zotti for providing the model count data which features in Figure \ref{fig:source_counts}, and Walter Brisken for his Cassbeam software which was used to generate the EVLA antenna beam patterns in Figure \ref{fig:beam_patterns}. We also thank the other participants of the Second Workshop on 3rd Generation Calibration in Radio Astronomy, where a discussion session inspired the simulations in the Appendix. We thank Vivek Dhawan for useful discussions regarding the EVLA pointing accuracy. Some of the figures in this paper were produced with APLpy, an open-source astronomical plotting package for Python ({\tt http://aplpy.github.com}). This research has made use of NASA's Astrophysics Data System. This paper is dedicated to Steve Rawlings.

\appendix
\section{Interpretation of the differential gain solutions}

\begin{figure*}
\begin{center}
\setlength{\unitlength}{1cm}
\begin{picture}(16,3.3)
\put(-1.6,-0.7){\includegraphics{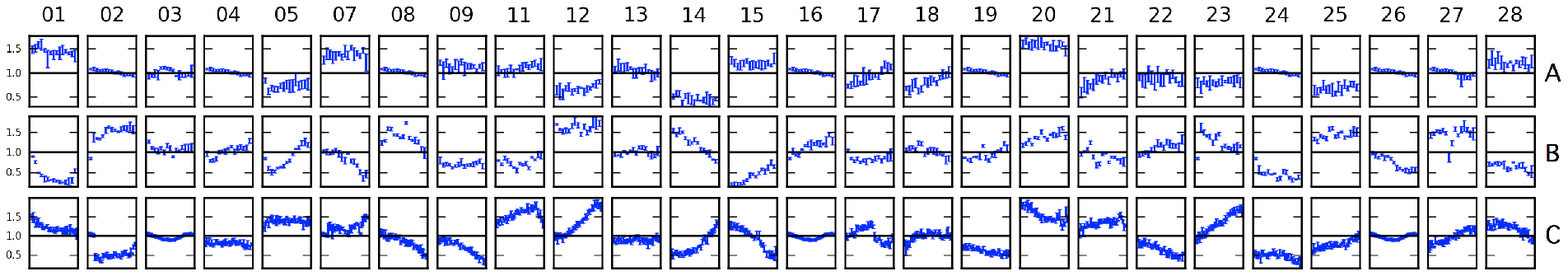}}
\end{picture}
\caption{\label{fig:whdf_dE} Mean amplitudes of the differential gain solutions for three different Measurement Sets (rows A, B and C). Each column corresponds to an antenna in the array.}
\end{center}
\end{figure*}

\begin{figure}
\vspace{160mm}
\nonumber
\centering
\includegraphics[width= \columnwidth]{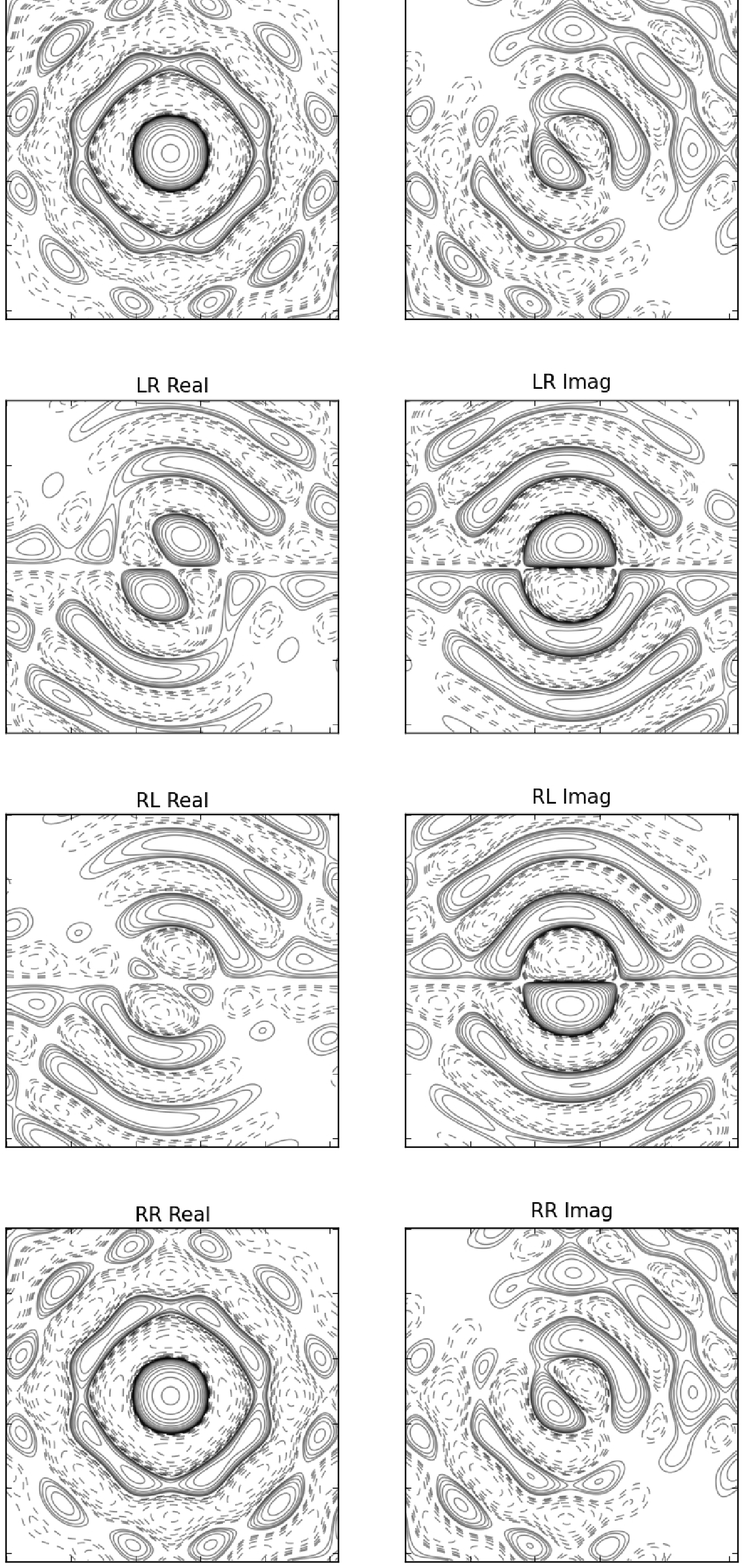} 
\caption{\label{fig:beam_patterns}Complex antenna beam patterns for all four polarizations products as generated by the Cassbeam software. The polarization product and whether the beam pattern corresponds to the real or imaginary component is indicated above each frame. These are rendered as FITS images and fed through a MeqTrees simulation module to investigate the effects of primary beam patterns on various source configurations. For the diagonal terms in the Jones matrix (LL and RR) the base contour level is $\pm$0.005, with adjacent positive and negative contours increasing in multiples of $\sqrt{2}$. The off-diagonal terms (RL and LR) have a base contour level of $\pm$0.0003 and the same increments as the diagonal terms. The images span a sky area of approximately 54 arcminutes.}
\end{figure}

Self-calibration of a radio interferometric data set involves the generation of a model visibility set from assumed models of the sky and the instrument. Model visibilities are usually predicted by evaluating the radio interferometer measurement equation (RIME; e.g.~Smirnov, 2011a). In the most commonly employed method of calibration the RIME has a solvable complex gain term per feed, per antenna, and a numerical algorithm is used to minimise the difference between the observed and the model visibility data. The best-fitting complex gain terms (which are functions of both time and frequency) are applied to correct the observed data and remove the instrumental gain drifts. These corrected data are then imaged for deconvolution or to refine the sky model for further calibration.

In the case where there is a single dominating source in the field (e.g.~a typical observation of a phase calibrator) both the observed and model visibility function, and the associated complex gain corrections, will be dominated by the contribution of this source. If the dominating source is away from the pointing centre (as is the case for the observations presented in this paper) then the time- and frequency-dependent behaviour of the complex gain solutions derived from self-calibration will be dominated by any instrumental or atmospheric effects which are \emph{specific to the direction towards that source}. In the case where the target field contains many strong sources, each of which is subject to a direction-dependent effect (DDE; e.g.~primary beam or ionospheric effects) the traditional approach of solving for a single complex gain term per antenna is insufficient if high fidelity or high dynamic range imaging is required.

Continuum imaging in the presence of sources which are blighted by strong DDEs results in maps whose dynamic range is limited by calibration artefacts rather than thermal noise, as is the ideal outcome. Self-calibration of the field will generate solutions dominated by the strongest source(s). Deriving a single complex gain correction is insufficient, and deficiencies in either the instrumental or sky model will manifest themselves as corrupted versions of the point spread function centred on the fainter sources, which deconvolution is unable to remove. 

The usual approach to mitigating these effects is to employ some form of peeling algorithm (e.g.~Noordam, 2004). This is an iterative process whereby the sources are treated in order of decreasing brightness. Self-calibration is performed on a per-source basis on a visibility set which is phase-rotated to the position of the source in question. The best fitting model is computed and subtracted from the original data. These residuals are then used as the starting point for the second brightest problem source, and the process repeats until an acceptable map is achieved. Aside from the fact that this process is somewhat unwieldy and prone to user error, it may have trouble converging in the scenario where there are multiple confusing sources of similar brightness.

A more flexible and generic alternative now exists in the form of the differential gains algorithm (Smirnov, 2011b). This avoids the generation of intermediate data products by forming simultaneous solutions for the complex receiver gains against an all-inclusive sky model on short timescales, as well as solving for additional gain terms on longer timescales for a subset of dominating sources. The solution interval for the second component should be matched to the maximum interval over which one expects the DDE to be roughly constant. This maximises the signal to noise in the measurement, minimises the degrees of freedom in the fit, and effectively time-smears out contributions from the other sources in the field. For source subtraction purposes (as implemented in this paper) the best fitting visibility models for each of the sources for which differential gain solutions are computed are subtracted from the data and the residual data are imaged. 

\subsection{Gain solutions}

The complex gain solutions themselves encode much information (Smirnov, 2011c). For the calibration process to be successful there is an implicit (and justifiable) assumption that (direction-dependent) gains vary smoothly with frequency and time. As mentioned in Section \ref{sec:subtraction} examining the solutions can provide a valuable diagnostic by highlighting deviations from smooth behaviour. The mean amplitudes of the differential gain solutions for 4C 00+02 for three different Measurement Sets (dubbed A, B and C) can be see in Figure \ref{fig:whdf_dE}. Each column corresponds to a single EVLA antenna. The plots show the mean amplitude of each solution as a function of time, averaged across the band. Rows A and B are derived from shorter 1.5 hour SBs and row C is derived from a 3.5 hour observation.

The smoothness of these solutions in time is immediately apparent. As severe as the presence of this confusing source was in terms of obtaining a scientifically useful map, the DDEs that it was subject to can be tracked by the solver, and despite the factor of $\sim$100 by which 4C 00+02 was attenuated by the primary beam response there is still ample signal to noise to form the solutions.

Discontinuities in the solutions (e.g.~27-B) are indicative of amplitude spikes on antennas that were missed during initial flagging. As the amplitude jumps suddenly the gain correction drops to compensate.

Can the origin of the observed temporal gain variations be deduced? The most significant causes of DDEs in interferometry data are the ionosphere and the primary beam response of the elements that make up the array. The ionosphere manifests itself as a dynamic phase screen over the array (Intema et al., 2009) and is a significant problem for observing with long baseline arrays at low frequencies, however the X-band observations presented in this paper are unlikely to be significantly affected by it. 

\subsection{The EVLA primary beam}

As mentioned in Section \ref{sec:subtraction} it is speculated that the gain drifts are caused by the a combination of the structure in the primary beam pattern and its apparent rotation on the sky as the observation progresses. Figure \ref{fig:beam_patterns} shows a simulated complex beam pattern for the EVLA at 8.4 GHz generated using the Cassbeam package (Walter Brisken, private communication). Cassbeam takes a parametrized model of the Cassegrain antenna optics (Brisken, 2003) and uses ray tracing to compute a 2$\times$2 Jones matrix (Smirnov, 2011a and references therein) describing the effect of the antenna on the incoming radiation as a function of direction. This complex-valued matrix is visualised in Figure \ref{fig:beam_patterns}.

The real components of the diagonal terms (LL and RR) show clearly the main lobe of the beam and the azimuthal structure beyond this induced by the antenna optics. The off-diagonal terms (LR and RL) effectively show the instrumental polarization of the EVLA, the so-called `beam squint' introduced by the fact that the two receptors sensitive to orthogonal polarization modes are not co-spatial. Polarization will not be discussed further.

Such beam models can also be used by the A-projection algorithm (Bhatnagar et al., 2008) both to predict model visibilities during calibration, and to apply a correction for known beam effects during imaging.

It has already been noted that the confusing source in the observations of the WHDF was situated close to the first sidelobe. Rotation of the beam on the sky will therefore cause the gain to drift according to the azimuthal asymmetries. This clearly does not tell the whole story as the gains in Figure \ref{fig:whdf_dE} do not exhibit similar behaviour from antenna to antenna. The hypothesised explanation for this is pointing error: radio telescopes have varying degrees of pointing accuracy which has the effect of shifting the beam patterns shown in Figure \ref{fig:beam_patterns} away from the nominal pointing centre. The chance positioning of the confusing source in a region of the beam that has azimuthal structure and steep radial gradients causes the gain to exhibit strong behaviour in time and frequency. The shifting of the beam pattern on the sky due to pointing error causes these variations to differ from antenna to antenna. The next step is to attempt a simulation which replicates the observations.

\subsection{Simulated observation}

The CASA sm tool was used to generate a Measurement Set with pointing direction and frequency range consistent with the real observations but with a six-hour track length, and antenna positions matching those of the the EVLA D-configuration with all 27 dishes. As per the averaged real data 8 $\times$ 32 MHz channels were used.

The Measurement Set was then filled with simulated visibilities using the Siamese framework within MeqTrees. The key part of this framework is the BeamSims module which is able to read the simulated beams presented above as gridded FITS images. The module performs interpolation of the beam patterns as well as sky rotation. Crucially, pointing errors can also be applied. Each antenna is assigned a random pointing offset in two orthogonal directions on the sky, with a value between 0 and 10 arcseconds, a conservative estimate of the true pointing accuracy of the EVLA (Vivek Dhawan, private communication). 

\begin{figure*}
\begin{center}
\setlength{\unitlength}{1cm}
\begin{picture}(16,4.2)
\put(-1.5,-1.0){\includegraphics{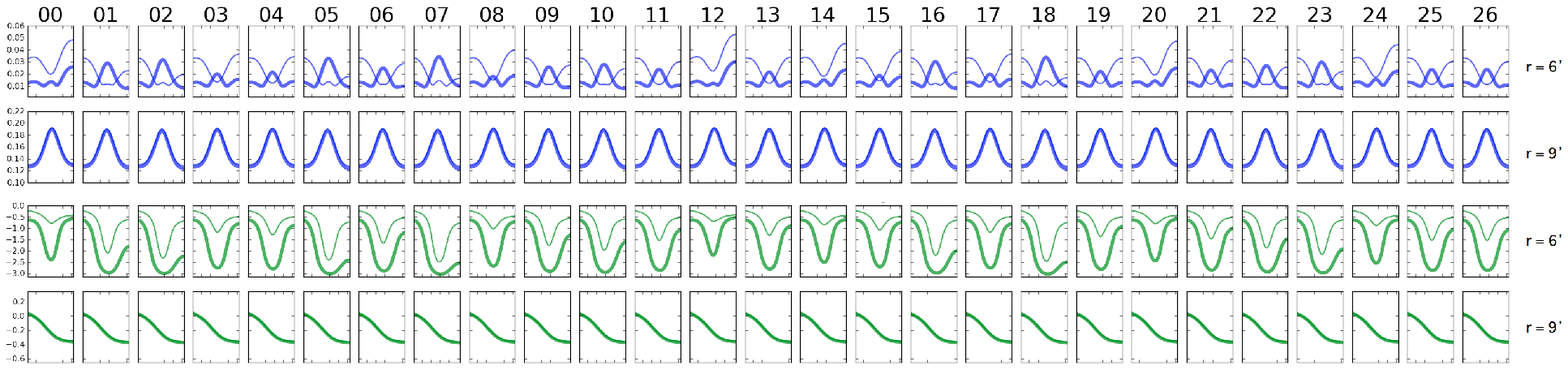}}
%\put(-1.7,-0.8){\special{psfile=EJones_sim_phase.eps vscale=92.0 hscale=92.0 voffset=0hoffset=0}}
\end{picture}
\caption{\label{fig:EJones} Simulated values of the complex beam gain (blue = amplitude, normalized to the maximal gain close to the beam centre; green = phase, in radians) as a function of time for each antenna in a six-hour EVLA simulation. Note that the phases in the r~=~9' case have been rotated by $\pi$ radians to make the plot clearer. A random pointing error with a maximum value of 10" is applied to each antenna, equivalent to shifting the beam patterns shown in Figure \ref{fig:beam_patterns} by a random amount in two orthogonal directions on the sky. For each complex component the plots show the behaviour for a source at a radius of 6 arcminutes from the phase centre (as per the EVLA observations of the WHDF) and at a radius of 9 arcminutes from the phase centre (corresponding to the approximate centre of the first sidelobe). The two lines in each panel correspond to two different channels in the simulation in order to show the strong chromatic effect in the r~=~6' case. The thick line is the 8.364 GHz and the thin line is 8.556 GHz, corresponding to the channels that span most of the bandwidth of the actual observations. The salient point that this plot makes is that a combination of pointing error, azimuthal structure in the beam and the chance positioning of a strong source in a part of the beam exhibiting strong radial gradients can result in beam gains with significantly different temporal and spectral behaviour from antenna to antenna.}
\end{center}
\end{figure*}

Two scenarios are now simulated. The first is with a source that is 6 arcminutes from the pointing centre, consistent with the location of 4C 00+02 in the WHDF observations. The second is a simulation of a source at a radius of 9 arcminutes, which places it in the approximate centre of the first sidelobe. As the visibilities are computed the complex values of the applied beam gain for each of these sources, per integration time, per channel, per antenna are exported from the RIME tree. Two channels corresponding to 8.364 and 8.556 GHz are selected and plotted as a function of time in Figure \ref{fig:EJones}. The upper two rows show the amplitude values and the lower two rows show the phases. The thickest line represents the lowest frequency channel. 

This simulation at least qualitatively reproduces the observed behaviour. For the source with a separation of 6 arcminutes there is large vertical scatter in the beam gains between antennas. There are periods where gains rise as others fall and the variation is strongly chromatic. By contrast the source at a radial separation of 9 arcminutes, although it exhibits temporal variation due to the rotation of the beam pattern, exhibits much less variation between antennas. The source at 6 arcminutes is in a part of the beam which has strong gain gradients, and the pointing error exacerbates the disparities between antennas.

\bsp % ``This paper has been produced using the ...''

\label{lastpage}


\begin{thebibliography}{}

\bibitem[Abazajian et al.(2009)]{abazajian2009} Abazajian, K.~N., Adelman-McCarthy, J.~K., Ag{\"u}eros, M.~A., et al.\ 2009, ApJ Suppl., 182, 543 

\bibitem[Antonucci(1993)]{1993ARA&A..31..473A} Antonucci, R.\ 1993, ARA\&A, 31, 473

\bibitem[Bhatnagar et al.(2008)]{bhatnagar2008} Bhatnagar, S., Cornwell, T.~J., Golap, K., \& Uson, J.~M.\ 2008, \aap, 487, 419

\bibitem[Baugh et al.(2005)]{baugh2005} Baugh C.~M., Lacey C.~G., Frenk C.~S., Granato G.~L., Silva L., Bressan A., 
Benson A.~J., Cole S., 2005, MNRAS, 356, 1191 

\bibitem[Bielby et al.(2012)]{bielby2012} Bielby, R.~M., Hill, M.~D., Metcalfe, N., \& Shanks, T.\ 2012, MNRAS, 419, 1315 

\bibitem[B{\"o}hm \& Ziegler(2007)]{bohm2007} B{\"o}hm, A., \& Ziegler, B.~L.\ 2007, ApJ, 668, 846 

\bibitem[Bonfield et al.(2011)]{bonfield2011} Bonfield D.~G., et al., 2011, MNRAS, 416, 13 

\bibitem[Brandt et al. (2001)]{brandt2001} Brandt W.~N., et al., 2001, AJ, 122, 2810

\bibitem[Brisken(2003)]{brisken2003} Brisken, W.\ 2003, \emph{Using GRASP8 to study the VLA primary beam}, EVLA Memo 58, {\tt http://www.aoc.nrao.edu/evla/memolist.shtml}

\bibitem[Cimatti et al.(2002)]{cimatti2002} Cimatti A., et al., 2002, A\&A, 391, L1 

\bibitem[Cirasuolo et al.(2010)]{2010MNRAS.401.1166C} Cirasuolo, M., 
McLure, R.~J., Dunlop, J.~S., et al.\ 2010, MNRAS, 401, 1166 

\bibitem[de Zotti et al.(2010)]{dezotti2010} de Zotti, G., Massardi, M., Negrello, M., \& Wall, J.\ 2010, A\&A Rev., 18, 1 
 
\bibitem[Downes et al.(1986)]{downes1986} Downes, A.~J.~B., Peacock, J.~A., Savage, A., \& Carrie, D.~R.\ 1986, MNRAS, 218, 31

\bibitem[Dunlop et al.(2003)]{2003MNRAS.340.1095D} Dunlop, J.~S., McLure, 
R.~J., Kukula, M.~J., et al.\ 2003, MNRAS, 340, 1095 

\bibitem[Fabian (1999)]{fabian1999} Fabian A.~C., 1999, MNRAS, 308, L39 

\bibitem[Fixsen et al.(2011)]{2011ApJ...734....5F} Fixsen, D.~J., Kogut, A., Levin, S., et al.\ 2011, ApJ, 734, 5 

\bibitem[Fomalont et al.(2002)]{fomalont2002} Fomalont, E.~B., Kellermann, K.~I., Partridge, R.~B., Windhorst, R.~A.,  \& Richards, E.~A.\ 2002, \aj, 123, 2402 

\bibitem[Greisen(2003)]{greisen2003} Greisen, E.~W.\ 2003, Information Handling in Astronomy - Historical Vistas, 285, 109 

\bibitem[Gunn \& Shanks(1999)]{gunn1999} Gunn K.~F. \& Shanks T., 1999, arXiv:astro-ph/9909089 

\bibitem[Gunn(1999)]{gunnphd} Gunn K.~F., 1999, PhD Thesis, Durham University  

\bibitem[Hatziminaoglou et al.(2010)]{hatz2010} Hatziminaoglou E., et al., 2010, A\&A, 518, L33

\bibitem[Henkel \& Partridge(2005)]{henkel2005} Henkel, B., \& Partridge, R.~B.\ 2005, \apj, 635, 950 

\bibitem[Heywood et al.(2011)]{heywood2011} Heywood, I., Armstrong, R.~P., Booth, R., et al.\ 2011, arXiv:1103.0862 

\bibitem[Hickox et al.(2012)]{hickox2012} Hickox R.~C., et al., 2012, MNRAS, 421, 284 

\bibitem[Hill \& Shanks(2011a)]{hill2011a} Hill M.~D., Shanks T., 2011, MNRAS, 410, 762 

\bibitem[Hill \& Shanks(2011b)]{hill2011b} Hill M.~D., Shanks T., 2011, MNRAS, 414, 1875 

\bibitem[Intema et al.(2009)]{intema2009} Intema, H.~T., van der Tol, S., Cotton, W.~D., et al.\ 2009, \aap, 501, 1185 

\bibitem[Ivison et al. (1998)]{ivison1998} Ivison R.~J., Smail I., Le
Borgne J.-F., Blain A.~W., Kneib J.-P., Bezecourt J., Kerr T.~H., Davies
J.~K., 1998, MNRAS, 298, 583

\bibitem[Ivison et al.(2002)]{ivison2002} Ivison, R.~J., Greve, 
T.~R., Smail, I., et al.\ 2002, \mnras, 337, 1

\bibitem[Jarvis \& Rawlings (2004)]{jarvis2004} Jarvis, M.~J., Rawlings S., 2004, NewAR, 48, 1173 

\bibitem[Jarvis et al.(2010)]{jarvis2010} Jarvis, M.~J., et al., 2010, MNRAS, 409, 92

\bibitem[Jarvis(2011)]{jarvis2011} Jarvis, M.~J.\ 2011, arXiv:1107.5165 

\bibitem[Kuraszkiewicz et al.(2003)]{kura2003} Kuraszkiewicz J.~K., et al., 2003, ApJ, 590, 128 

\bibitem[Lutz et al.(2010)]{lutz2010} Lutz D., et al., 2010, ApJ, 712, 1287

\bibitem[Maddox et al.(2012)]{maddox2012} Maddox N., Hewett P.~C., P{\'e}roux C., Nestor D.~B., Wisotzki L., 2012, MNRAS, 424, 2876

\bibitem[Mart{\'{\i}}nez-Sansigre et al.(2005)]{martinez-sansigre2005} Mart{\'{\i}}nez-Sansigre, A., Rawlings, S., Lacy, M., et al.\ 2005, Nature, 436, 666 

\bibitem[McCracken et al.(2000)]{mccracken2000} McCracken, H.~J., 
Shanks, T., Metcalfe, N., Fong, R., \& Campos, A.\ 2000, \mnras, 318, 913 

\bibitem[Metcalfe et al.(1991)]{metcalfe1991} Metcalfe, N., Shanks, 
T., Fong, R., \& Jones, L.~R.\ 1991, MNRAS, 249, 498 

\bibitem[Metcalfe et al.(1995)]{metcalfe1995} Metcalfe, N., Shanks, 
T., Fong, R., \& Roche, N.\ 1995, MNRAS, 273, 257 

\bibitem[Metcalfe et al.(2001)]{metcalfe2001} Metcalfe, N., Shanks, 
T., Campos, A., McCracken, H.~J., \& Fong, R.\ 2001, \mnras, 323, 795 

\bibitem[Metcalfe et al.(2006)]{metcalfe2006} Metcalfe, N., Shanks, 
T., Weilbacher, P.~M., et al.\ 2006, \mnras, 370, 1257

\bibitem[Molinari et al.(2011)]{molinari2011} Molinari S., et al., 2011, ApJ, 735, L33 

\bibitem[Noordam(2004)]{noordam2004} Noordam, J.~E.\ 2004, Proceedings of the SPIE, 5489, 817 

\bibitem[Noordam \& Smirnov(2010)]{noordam2010} Noordam, J.~E., \& Smirnov, O.~M.\ 2010, A\&A, 524, A61 

\bibitem[Padovani et al.(2009)]{padovani2009} Padovani, P., Mainieri, V., Tozzi, P., et al.\ 2009, ApJ, 694, 235 

\bibitem[Page et al.(2004)]{page2004} Page, M.~J., Stevens, J.~A., Ivison, R.~J., \& Carrera, F.~J.\ 2004, ApJ Letters, 611, L85 

\bibitem[Page et al.(2012)]{page2012} Page M.~J., et al., 2012, Natur, 485, 213

\bibitem[Polletta et al.(2000)]{polletta2000} Polletta M., Courvoisier T.~J.-L., Hooper E.~J., Wilkes B.~J., 2000, A\&A, 362, 75 

\bibitem[Sanders et al.(1988)]{sanders1988} Sanders D.~B., Soifer B.~T., Elias J.~H., 
Madore B.~F., Matthews K., Neugebauer G., Scoville N.~Z., 1988, ApJ, 325, 74 

\bibitem[Seiffert et al.(2011)]{seiffert2011} Seiffert, M., Fixsen, D.~J., Kogut, A., et al.\ 2011, ApJ, 734, 6 

\bibitem[Serjeant et al.(2010)]{serjeant2010} Serjeant S., et al., 2010, A\&A, 518, L7

\bibitem[Shi et al.(2005)]{shi2005} Shi Y., et al., 2005, ApJ, 629, 88 

\bibitem[Simpson et al.(2006)]{simpson2006} Simpson, C., 
Mart{\'{\i}}nez-Sansigre, A., Rawlings, S., et al.\ 2006, MNRAS, 372, 741 

\bibitem[Simpson et al.(2012)]{simpson_c2012} Simpson C., et al., 2012, MNRAS, 421, 3060 


\bibitem[Simpson et al.(2012)]{simpson2012} Simpson J.~M., et al., 2012, arXiv, 
arXiv:1206.4692 

\bibitem[Smirnov(2011)]{smirnov2011a} Smirnov, O.~M.\ 2011a, A\&A, 527, A106 

\bibitem[Smirnov(2011)]{smirnov2011b} Smirnov, O.~M.\ 2011b, A\&A, 527, A107 

\bibitem[Smirnov(2011)]{smirnov2011c} Smirnov, O.~M.\ 2011c, A\&A, 527, A108 

\bibitem[Smol{\v c}i{\'c} et al.(2009)]{smolcic2009} Smol{\v c}i{\'c} V., et al., 2009, ApJ, 
696, 24 

% \bibitem[Thompson et al.(2001)]{thompson2001} Thompson, A.~R., Moran, J.~M., \& Swenson, G.~W., Jr.\ 2001, ``Interferometry and synthesis in radio astronomy"~2nd ed.~ New York : Wiley, ISBN 0471254924 

\bibitem[Vallbe-Mumbru.(2004)]{vallbe-mumbru2004} Vallbe-Mumbru M., 2004, PhD thesis, Durham University

\bibitem[Vernstrom et al.(2011)]{vernstrom2011} Vernstrom, T., Scott, D., \& Wall, J.~V.\ 2011, MNRAS, 415, 3641 

\bibitem[Wardlow et al.(2011)]{wardlow2011} Wardlow J.~L., et al., 2011, MNRAS, 415,  1479 

\bibitem[Wilman et al.(2008)]{wilman2008} Wilman, R.~J., Miller, L., Jarvis, M.~J., et al.\ 2008, MNRAS, 388, 1335 

\bibitem[Windhorst et al.(1993)]{windhorst1993} Windhorst, R.~A., Fomalont, E.~B., Partridge, R.~B., \& Lowenthal, J.~D.\ 1993, ApJ, 405, 498 

\end{thebibliography}
\end{document}